\documentclass[%
 reprint,showkeys,onecolumn,
 amsmath,amssymb,
 aps,
]{revtex4-2}
\usepackage{xcolor}
\usepackage{graphicx}
\usepackage{dcolumn}
\usepackage{bm}

\begin{document}

\preprint{APS/123-QED}

\title{Casimir Energy for Lorentz-Violating Scalar Field with Helical Boundary Condition in $d$ Spatial Dimensions}

\author{M. A. Valuyan}
 \altaffiliation[Also at ]{Energy and Sustainable Development Research Center, Semnan Branch, Islamic Azad University, Semnan, Iran}
\email{m-valuyan@sbu.ac.ir;ma.valouyan@iau.ac.ir}
\affiliation{%
 Department of Physics, Semnan Branch, Islamic Azad University, Semnan, Iran.
}%




\date{\today}

\begin{abstract}
Delving into spring-like helical configurations, such as DNA within our cells, motivates physicists to inquire about the effects of such structures in the realm of quantum field theory, specifically unraveling their manifestation of effectiveness in Casimir energy. To explore this, we initiated our investigation with the Casimir effect and proceeded to calculate the Casimir energy, along with its thermal and radiative corrections, for massive and massless Lorentz-violating scalar fields across $d+1$ dimensional space-time. Adhering to the principle that the renormalization program should be consistent with the boundary conditions applied to the quantum fields was paramount. Accordingly, one-loop correction to the Casimir energy for both massive and massless scalar fields under helical boundary conditions by employing a position-dependent counterterm was performed. Our results, spanning across spatial dimensions, exhibited convergence and coherence with established physical grounds. Additionally, we presented the Casimir energy density using graphical plots for systems featuring time-like and space-like Lorentz violations across all spatial dimensions, followed by a comprehensive discussion of the obtained results.
\end{abstract}

\maketitle

\section{\label{Intro.}Introduction}
H.B.G. Casimir’s seminal prediction of the Casimir effect, made over 60 years ago, set the stage for extensive research in this field\,\cite{h.b.g.}. The Casimir effect, which arises from distortions in the vacuum, has been the subject of thorough investigation over the years, with notable contributions from various works\,\cite{other.works.1,other.works.2,all.method.ZO}. Such distortions in the vacuum can be induced by the presence of boundaries in the space-time or by external factors such as gravity. With significant advancements in experimental accuracy, research interest in the Casimir effect has surged, as evidenced by the growing attention it has garnered over time\,\cite{experiment.accuracy.1,experiment.accuracy.2}. This heightened interest is not only driven by the theoretical and mathematical aspects of the Casimir effect but also by its wide-ranging and impactful applications, given its characteristic as a quantum effect with observable macroscopic implications\,\cite{NEW.DEVELOP.}. Hence, the Casimir energy and force between two objects like as two points, two parallel plates or two surfaces were investigated for multiple quantum fields\,\cite{Ddim.1,Ddim.2}. With advances in technology at the micro and nano scales, the importance of Casimir's effect in understanding and predicting various phenomena has increased. In biophysics, for example, as devices approach the size of a single cell, it is possible to describe how cells attach to each other by Casimir force or the like. The folding of proteins and some other cellular interactions can also be explained by the Casimir effect \cite{bio.Casimir.1,bio.Casimir.2,bio.Casimir.3}. The interplay of geometry, boundary conditions, and temperature is crucial in understanding the behavior of the Casimir force on material components, especially in the context of nanoelectromechanical and microelectromechanical systems \cite{Thermal.Cas.2,Thermal.Cas.3,Thermal.Cas.4}.
\par The preservation of Lorentz symmetry is a fundamental assumption in most traditional studies of the Casimir effect. However, at high energy scales, certain theoretical frameworks fail to uphold this symmetry, leading to anisotropic spacetime, including anisotropy in the time dimension. In such scenarios, the directions where Lorentz symmetry is violated are typically specified by a constant unit 4-vector. An immediate consequence of this anisotropic spacetime is the modification of the spectrum of the Hamiltonian operator and, consequently, the dispersion relation. This, in turn, alters the energy modes of the quantum field, resulting in modifications to the Casimir energy density. As a result, the violation of Lorentz symmetry has garnered significant attention from physicists investigating Casimir effect problems in recent years\,\cite{magnetic.}. The early studies exploring Casimir energy in the context of Lorentz-violating quantum fields were documented in Ref. \cite{15ta17}, with further analysis of the Casimir effect for the H\v{o}rava-Lifshitz-like massless scalar field presented in \cite{18} and \cite{19}. Additionally, the Casimir effect for the massive scalar field was investigated in \cite{cruz.1,cruz.2}. String theory, typically associated with high-energy physics, has also been a subject of interest in the context of Lorentz symmetry breaking, as highlighted in Ref.  \cite{15}. Furthermore, investigations into the effects of Lorentz violation on the Casimir energy at low-energy scales have been conducted in Refs. \cite{16ta24.1,16ta24.2}.
\par
The radiative correction to the Casimir energy has emerged as a significant area of interest among physicists, garnering considerable attention,\cite{Bordag.1,Bordag.2}. Various methods for renormalization and regularization have been devised to facilitate accurate calculations in this domain. These methods encompass the Zeta function regularization technique, \cite{other.regular.tech.1,other.regular.tech.3,other.regular.tech.4}, Heat kernel series,\cite{Heat.kernel}, and Green's function method,\cite{BSS.2,BSS.1}. In this regard, comprehensive investigations have been undertaken to delve into radiative corrections to the Casimir energy across various fields and geometries \cite{developement.cas.,NEW.DEVELOP.}. It is imperative to note that in scenarios where non-trivial boundary conditions or topological factors influence the quantum field, all components of the renormalization program, including counterterms, must align consistently with these conditions. The selection of appropriate counterterms holds particular significance as they play a crucial role in renormalizing the bare parameters within the Lagrangian. Incorrect choices may result in the incomplete removal of divergences, thereby yielding divergent values for some physical quantities \cite{CAVALCANTI.1,CAVALCANTI.2,CAVALCANTI.3}. In line with these considerations, a systematic renormalization program was proposed by S.S. Gousheh et al., which facilitates the derivation of counterterms consistent with boundary conditions,\cite{EUR-Reza,JHEP.REZA}. Notably, the counterterms obtained within their renormalization program are position-dependent. Comprehensive details regarding their renormalization program, including the derivation of counterterms from $n$-point functions within renormalized perturbation theory, have been elucidated in our previously published works,\cite{RC.MAN.EPJP,LP.MAN}. Our present study utilizes the results for Lorentz violating scalar field to automatically extract counterterms from the renormalization program, allowing us to obtain the vacuum energy of the system up to the first order of the coupling constant $\lambda$. Furthermore, the helical boundary condition was chosen for our problem, inspired by the helix-shaped structures in nature, such as DNA and cell membrane proteins. These biological entities have sizes ranging from nanometers to micrometers, resulting in significant Casimir energy and force. This significant value for the Casimir energy reinforces the motivation to explore the Casimir energy in helix-like structures across different quantum fields. Additionally, considering the importance of temperature in biological systems, our study not only calculates the leading-order Casimir energy but also discusses the thermal corrections to the Casimir energy. It is important to note that the Casimir energy and its radiative correction for a massive and massless scalar field obeying helical boundary conditions in $3+1$ dimensions were documented in Ref. \cite{failing.LV}. In this context, we generalize our investigation to compute the leading order and its thermal correction to the Casimir energy for Lorentz-violated massive/massless scalar fields with helical boundary conditions in $d+1$ space-time dimensions. Furthermore, we delve into computing the radiative corrections to the Casimir energy for self-interacting Lorentz-violated massive/massless scalar field with helical boundary conditions in $d+1$ space-time dimensions. Another distinction in our computation of the radiative correction to the Casimir energy compared to those documented in Ref. \cite{failing.LV} lies in the type of counterterm utilized. We employ the self-consistent renormalization program, which enables the derivation of counterterms consistent with the imposed boundary conditions, leading to counterterms that are position-dependent. All obtained results are consistent with expected physical grounds. The structure of this paper is as follows: In Section \ref{Sec:model}, we introduce a model for obtaining the dispersion relations for categorized directions of Lorentz violation. Then, a model for executing the renormalization program, including the position-dependent counterterm to compute the radiative correction term and deduce the vacuum energy, is presented. Section \ref{sec.zero.order} and its subsections detail our calculations to obtain the leading order Casimir energy for Lorentz-violated massive and massless free scalar fields. The next section presents the computation of the thermal correction to the Casimir energy, followed by Section \ref{Sec:RC}, which details the computation of the one-loop correction to the Casimir energy for the self-interacting massive/massless Lorentz-violating scalar field in d spatial dimensions ($\mathcal{O}(\lambda)$). The paper concludes in Section \ref{sec:conclusion}, summarizing the main results.

\section{The model}\label{Sec:model}
In this section, a model to compute the admissible modes for the Lorentz violating scalar field was briefly presented. Thus, starting from the Lagrangian of the free real scalar field with the Lorentz violating term, we have:
\begin{eqnarray}\label{lagrangy}
        \mathcal{L}=\frac{1}{2}[\partial_\mu\phi(x)\partial^\mu\phi(x)
        +\alpha(u.\partial\phi(x))^2]-\frac{1}{2}m^2\phi^2(x),
\end{eqnarray}
where the parameter $m$ is the mass of the quantum field, and the dimensionless coefficient $\alpha\ll1$ shows the scale of the Lorentz symmetry breaking. This parameter encodes the Lorentz violation by multiplying the scalar field derivative with a vector $\mathbf{u}=(u_0,u_1,u_2,u_3,...,u_d)$ which determines the direction of Lorentz violation in space-time\,\cite{cruz.1,cruz.2}. The vector $\mathbf{u}$ indicates that there are $d+1$ possible directions for Lorentz symmetry breaking.  We have classified these symmetry-breaking directions into four distinct cases. The Time-Like(TL) Lorentz violating scalar field occurs when the vector $\mathbf{u}=(1,0,0,...,0)$. For the vector $\mathbf{u}=(0,1,0,0,...,0)$, one of space-like Lorentz violating direction occurs, which from now on we want to call this direction of symmetry breaking by $Lv_1$. For the vector $\mathbf{u}=(0,0,1,0,...,0)$, the Lorentz symmetry breaking in direction $x_2$ are occurred, and for this direction we choose the name $Lv_2$. Likewise, for other cases of Lorentz symmetry breaking in which the component of vector $\mathbf{u}$ is $u_{i\neq0,1,2}=1$, the name $Lv_s$ is selected. To trace the effects of Lorentz violation in quantum fields, it is essential to begin with the equation of motion. Thus, from the Lagrangian presented in Eq. (\ref{lagrangy}), we derive the following equation of motion:
\begin{eqnarray}\label{Eq.motion.}
  [\square+\alpha(\mathbf{u}.\partial)^2-m^2]\phi=0.
\end{eqnarray}
In $d+1$ dimensional space-time, we consider the following boundary condition to emulate a helical structure:
\begin{eqnarray}\label{Helix.BC.}
      \phi(t,x_1+a,x_2,...,x_d)=\phi(t,x_1,x_2+h,...,x_d).
\end{eqnarray}
Here, $h$ represents the pitch of the helix, and we refer to this condition as the ``helix boundary condition.'' For the case of TL Lorentz violating scalar field, the solution of Eq.\,(\ref{Eq.motion.}) after applying the helix boundary condition gives the following form of the wave-number:
\begin{eqnarray}\label{Wave.number.TL}
      (1+\alpha)\omega_n^2=k_1^2+k_2^2+k_3^2+...+k_d^2+m^2,
\end{eqnarray}
where the wavevector $k_1=\frac{2n\pi}{a}+k_2 r$, the parameters $r=\frac{h}{a}$ and $n=0,\pm1,\pm2,...$. The dispersion relation for the $Lv_1$ case yields the wave number as follows:
\begin{eqnarray}\label{Wave.number.SLx}
       \omega_n^2=(1-\alpha)k_1^2+k_2^2+k_3^2+...+k_d^2+m^2.
\end{eqnarray}
On the same way, for each of the other space-like Lorentz violating directions (such as directions $x_i$ with $i>1$), the corresponding wavevector $k_i^2$ on the right-hand side of Eq. (\ref{Wave.number.SLx}) transforms to $(1-\alpha)k_i^2$.
\par
The zero-point vacuum energy, along with its thermal and radiative corrections, is typically expressed as follows:
\begin{eqnarray}\label{Zero.point&Thermal term.1}
 E_{\mbox{\tiny Vac.}}=E^{(0)}+E^{(T)}+E^{(R)}=\frac{1}{2}\sum\omega
 +\sum\frac{\omega}{ e^{\beta(\omega-\mu)}-1}+E^{(R)}.
\end{eqnarray}
In the given equation, the first term, $E^{(0)}$, signifies the zero-point energy, while the second and third terms, $E^{(T)}$ and $E^{(R)}$, represent the thermal and radiative corrections of the vacuum energy, respectively. In the second term on the right-hand side of the above equation, the parameter $\mu$ is known as the chemical potential. Additionally, the parameter $\beta$, is defined as $\beta=T^{-1}$, with $T$ denotes the temperature. Following the initial definition of the Casimir energy, as documented in numerous prior studies, we have:
\begin{eqnarray}\label{cas.define.}
      E_{\mbox{\tiny Cas.}}=E^{(0)}_{\mbox{\tiny Cas.}}+E^{(T)}_{\mbox{\tiny Cas.}}+E^{(R)}_{\mbox{\tiny Cas.}}=\Big(E^{(0)}_{\mbox{\tiny B.C.}}-E^{(0)}_{\mbox{\tiny free}}\Big)+\Big(E^{(T)}_{\mbox{\tiny B.C.}}-E^{(T)}_{\mbox{\tiny free}}\Big)+\Big(E^{(R)}_{\mbox{\tiny B.C.}}-E^{(R)}_{\mbox{\tiny free}}\Big),
\end{eqnarray}
where $E^{(0)}_{\mbox{\tiny B.C.}}$ represents the vacuum energy accounting for the influence of boundary conditions, and $E^{(0)}_{\mbox{\tiny free}}$ denotes the vacuum energy in the absence of any boundary conditions. The term within the second parentheses illustrates the subtraction of the thermal correction to the vacuum energy with and without boundary conditions. In the subsequent section, we compute the leading-order Casimir energy $E^{(0)}_{\mbox{\tiny Cas.}}$, followed by the calculation of the thermal correction term to the Casimir energy in Section \ref{Sec:TC}. Likewise, the third term on the right-hand side of Eq. (\ref{cas.define.}) corresponds to the radiative correction to the Casimir energy. To elucidate the magnitude of radiative corrections to the Casimir energy, it is imperative to establish a theoretical framework accommodating self-interacting and Lorentz-violating massive/massless scalar fields. Consequently, we introduced a self-interacting term into the Lagrangian defined in Eq.\,\eqref{lagrangy}. Subsequently, we obtained:
\begin{eqnarray}\label{lagrangian.bare}
   \mathcal{L}=\frac{1}{2}[\partial_\mu\phi(x)\partial^\mu\phi(x)
        +\alpha(u.\partial\phi(x))^2]-\frac{1}{2}m_0^2\phi^2(x)
        -\frac{\lambda_0}{4!}\phi(x)^4,
\end{eqnarray}
where $\lambda_0$ and $m_0$ are the bare coupling constant and bare mass of the field, respectively. At the level of radiative corrections to the Casimir energy, it is essential to renormalize all bare parameters of the Lagrangian, such as mass $m_0$ and coupling constant $\lambda_0$. Typically, the role of counterterms in the renormalization program is to eliminate divergences arising from these bare parameters. The selection of an appropriate counterterm has been a subject of extensive debate within the literature \cite{EUR-Reza}. In the majority of prior studies, the free counterterm, designed for Minkowski space, has been utilized in renormalization programs \cite{free.counterterms.1,free.counterterms.2,free.counterterms.3}. Nevertheless, the importance of adopting a counterterm that is in line with the boundary conditions imposed was underscored in the discussions presented in Refs.\,\cite{JHEP.REZA}. The counterterms proposed in these references are notably position-dependent, unlike the free counterterms, thereby accounting for the influence of boundary conditions or backgrounds within the renormalization program. These  counterterms enable a self-consistent renormalization program for the Lagrangian’s bare parameters. In this study, we not only validate the efficacy of using this renormalization program but also employ it to determine the vacuum energy. The complete renormalization program and the derivation of the general form of the first-order vacuum energy\,($\mathcal{O}(\lambda)$) have been documented in our prior work \cite{LP.MAN}. Therefore, to avoid reiterating the detailed process, we will only utilize the established final form of the first-order vacuum energy. Consequently, we have:
\begin{eqnarray}\label{First.Order.Vacuum.En.}
E^{(R)}=\frac{-\lambda}{8}\int_V  G^2(x,x)d^{d}x,
\end{eqnarray}
where $G(x,x')$ is the Green's function. The final expression of the Green's function for the real scalar field bounded by the helical boundary condition in an arbitrary spatial dimension $d$ after Wick rotation becomes:
\begin{eqnarray}\label{Greens.Function.TLHBC}
    G(x,x')=\frac{1}{a}\int\frac{d\omega}{2\pi}
    \int\frac{\prod_{i=2}^{d-1}dk_i}{(2\pi)^{d-1}}
    \sum_{n=-\infty}^{\infty}\frac{e^{-\omega(t-t')}
    e^{ik_1(x_1-x'_1)}...e^{ik_d(x_d-x'_d)}}
    {\omega^2+k_1^2+k_2^2+...+k_d^2+m^2}.
\end{eqnarray}
The expression for Green’s function presented in the preceding equation pertains to a scalar field, wherein Lorentz symmetry is conserved. To obtain the Green's function expression for the case of TL Lorentz violating scalar field it suffices to transform the $\omega^2$ in the denominator of Eq. (\ref{Greens.Function.TLHBC}) to $(1+\alpha)\omega^2$. Similarly, for each of the space-like Lorentz violating directions (\emph{e.g.} directions $x_i$ with $i\geq1$), the corresponding wavevector $k_i^2$ in the denominator of Eq. (\ref{Greens.Function.TLHBC}) transforms to $(1-\alpha)k_i^2$. By substituting the obtained Green's function for each case of Lorentz symmetry breaking into Eq. (\ref{First.Order.Vacuum.En.}), and employing the proper regularization technique, we can compute the radiative correction to the Casimir energy. This step in the calculations will be presented in Section \ref{Sec:RC}.
\section{Zero-Order Casimir Energy}\label{sec.zero.order}
Using Eq. (\ref{cas.define.}), the total Casimir energy expression is formally given by
\begin{eqnarray}\label{Cas.En.Formal.Form}
  E^{(0)}_{\mbox{\tiny Cas.}} &=& \frac{1}{2}\sum_{n,k} ( \omega^{\mbox{\tiny Bound}}_n-\omega^{\mbox{\tiny Free}}_k),
\end{eqnarray}
where $\omega_k^{\mbox{\tiny Bound}}$ denotes the wave-number satisfying the boundary condition. Consequently, $\omega_k^{\mbox{\tiny Free}}$ denotes the wave-number value in free space-time, characterized by the absence of any boundary conditions. Owing to the multiple directions of Lorentz symmetry breaking, the calculation of the Casimir energy for the scalar field varies with each specific direction of violation. Consequently, to elucidate the intricacies of the calculations, the content has been subdivided into two subsections. Subsection \ref{subsec.ZeroEn.TL&Ls} details the calculations required to ascertain the Casimir energy related to the TL and $Lv_s$ directions of Lorentz symmetry breaking. Subsection \ref{subsec.ZeroEn.Lv1} will present the computation of the Casimir energy in the context of $Lv_1$ and $Lv_2$ Lorentz violations.
\subsection{The Case of TL and $Lv_s$ of Lorentz Violations}\label{subsec.ZeroEn.TL&Ls}
Using Eqs. (\ref{Wave.number.TL},\ref{Wave.number.SLx},\ref{Cas.En.Formal.Form}), for the case of TL/$Lv_s$ Lorentz violating massive scalar field the zero-order Casimir energy density is converted to:
\begin{eqnarray}\label{Zero.Cas.En.1}
      \mathcal{E}^{(0)}_{\mbox{\tiny Cas.}} &=&\frac{1}{2\sqrt{1+s\alpha}}\Bigg\{\int
      \frac{\prod_{i=2}^{d}dk_i}{(2\pi)^{d-1}a}
      \sum_{n=-\infty}^{\infty}\big[k_1^2+k_2^2+k_3^2+...+k_d^2+m^2\big]
      ^{1/2}-\int_{0}^{\infty}
      \frac{\Omega_{d}\kappa^{d-1}d\kappa}{(2\pi)^{d}}
      \big[\kappa^2+m^2\big]^{1/2} \Bigg\},
\end{eqnarray}
where $\kappa=(\kappa_1,\kappa_2,...,\kappa_d)$ and $s=\pm1$. Also $\Omega_d=\frac{2\pi^{d/2}}{\Gamma(d/2)}$ is the spatial angle in $d$ spatial dimensions. For the parameter $s=1$ the Casimir energy for the case of TL would be obtained and $s=-1$ refers to the case $Lv_s$ of the Lorentz symmetry breaking. By replacing the wavevector $k_1=\frac{2n\pi}{a}+k_2 r$ and employing changing of variables for integrals we obtain,
\begin{eqnarray}\label{Zero.Cas.En.2}
       \mathcal{E}^{(0)}_{\mbox{\tiny Cas.}} &=&\frac{1}{2a(2\pi)^{d-1}
       \gamma^{\frac{d+1}{2}}\sqrt{1+s\alpha}}
       \int\prod_{i=2}^{d}dz_i
      \sum_{n=-\infty}^{\infty}
      \bigg[\Big(\frac{2n\pi}{a}\Big)^2+z_2^2+...+z_d^2+M^2\bigg]^{1/2}\nonumber\\
      &&-\frac{\Omega_{d}m^{d+1}}{2(2\pi)^{d}\sqrt{1+s\alpha}}\int_{0}^{\infty}
      \xi^{d-1}\sqrt{\xi^2+1}d\xi,
\end{eqnarray}
where the parameter $\gamma=1+r^2$, $r=h/a$ and $M=m\sqrt{\gamma}$. Furthermore, the second expression on the right-hand side of the aforementioned equation involves the substitution of the variable $\xi=\kappa/m$. The summation over $n$ in Eq. (\ref{Zero.Cas.En.2}) results in divergence. In order to address this divergence, a regularization process is necessary. Therefore, the Abel-Plana Summation Formula (APSF), as stated below, was employed to regulate the summation form in Eq. (\ref{Zero.Cas.En.2}):
\begin{eqnarray}\label{APSF}
         \sum_{n=-\infty}^{\infty}\mathcal{F}(n)=2\int_{0}^{\infty}
         \mathcal{F}(x)dx+2i\int\frac{\mathcal{F}(it)-\mathcal{F}(-it)}{e^
         {2\pi t}-1}dt.
\end{eqnarray}
The first term on the right-hand side of Eq. (\ref{APSF}) is commonly divergent and referred to as the \emph{Integral term}. While the second term has usually a convergent contribution and is  known as the \emph{Branch-cut term}. It is important to note that the expression for APSF (as given in Eq. (\ref{APSF})) is valid only when the summand is even. In the current context, this condition holds true. Using Eq. (\ref{APSF}) for Eq. (\ref{Zero.Cas.En.2}) leads to
\begin{eqnarray}\label{Zero.Cas.En.3}
     \mathcal{E}^{(0)}_{\mbox{\tiny Cas.}} &=&\frac{2}{2(2\pi)^{d}
       \gamma^{\frac{d+1}{2}}\sqrt{1+s\alpha}}
       \int_{0}^{\infty}dX\int_{-\infty}^{\infty}
       \prod_{i=2}^{d}dz_i\Big(X^2+z_2^2+...+z_d^2+M^2\Big)^{1/2}+\mathcal{B}(a)
      \nonumber\\
      &&-\frac{\Omega_{d}m^{d+1}}{2(2\pi)^{d}\sqrt{1+s\alpha}}\int_{0}^{\infty}
      \xi^{d-1}\sqrt{\xi^2+1}d\xi,
\end{eqnarray}
where $\mathcal{B}(a)$ represent the branch-cut term. The divergent part arising from the first part of Eq. (\ref{Zero.Cas.En.3}) cancels out the divergent term contributed by the free space. Thus, the only remained term from Eq. (\ref{Zero.Cas.En.3}) would be
\begin{eqnarray}\label{ZO.Branch.cut.1}
     \mathcal{E}^{(0)}_{\mbox{\tiny Cas.}} &=&\frac{2i}{2(2\pi)^{d-1}a\gamma^{\frac{d+1}{2}}
     \sqrt{1+s\alpha}}\int_{0}^{\infty}dt\int_{-\infty}^{\infty}
     \prod_{i=2}^{d} dz_i\frac{\mathcal{Z}(it)-\mathcal{Z}(-it)}{e^{2\pi t}-1},
\end{eqnarray}
where
\begin{eqnarray}\label{Z(t).TL&Lvs}
\mathcal{Z}(x)=\left[\left(\frac{2\pi x}{a}\right)^2+z_2^2+...+z_d^2+M^2\right]^{1/2}.
\end{eqnarray}
Upon changing the variable $T=2\pi t/a$, setting $p^2=z^2+M^2$, denoting $z$ as the vector $z=(z_2,z_3,...,z_d)$, and changing the order of integration over $T$ and $z$, Eq. (\ref{ZO.Branch.cut.1}) is transformed as follows:
\begin{eqnarray}\label{ZO.Branch.cut.2}
\mathcal{E}^{(0)}_{\mbox{\tiny Cas.}} &=&\frac{-4\Omega_{d-1}}{2(2\pi)^{d}a\gamma^{\frac{d+1}{2}}
     \sqrt{1+s\alpha}}\int_{M}^{\infty}\frac{dT}{e^{2Ta}-1}
     \int_{0}^{\sqrt{T^2-M^2}}
     dpp^{d-2}\left[T^2-p^2-M^2\right]^{1/2}.
\end{eqnarray}
Considering $\frac{1}{e^{aT}-1}=\sum_{j=1}^{\infty}e^{-jaT}$ and computing the integrations written in aforementioned equation, the Casimir energy density for the case of TL/$Lv_s$ Lorentz violating scalar field in $d+1$ dimensions is obtained as:
\begin{eqnarray}\label{Zero.Cas.En.4}
     \mathcal{E}^{(0)}_{\mbox{\tiny Cas.}} &=&\frac{-2}{\sqrt{1+s\alpha}}\left(\frac{M}{2\pi a\gamma}\right)
     ^{\frac{d+1}{2}}\sum_{j=1}^{\infty}
     \frac{K_{\frac{d+1}{2}}(Maj)}{j^{\frac{d+1}{2}}}.
\end{eqnarray}
The leading-order Casimir energy for a massive scalar field constrained by helical boundary conditions in three spatial dimensions, which preserves Lorentz symmetry, was previously documented in Ref. \cite{Thermal.Cas.1}. Our derived result, as presented in Eq. (\ref{Zero.Cas.En.4}), aligns with the reported findings in proper limits ($\alpha\to0$). In order to compute the Casimir energy for the massless scalar field, we go back to Eq. (\ref{Zero.Cas.En.1}) and set the mass $m=0$. By repeating the aforementioned calculation process the result is obtained as:
\begin{eqnarray}\label{massless.Zero.Cas.TL.Lvs.}
     \mathcal{E}^{(0)}_{\mbox{\tiny Cas.}} \buildrel\mbox{\hspace{0.2cm}$m\to0$\hspace{0.2cm}}
     \over\longrightarrow-\frac{1}{\sqrt{1+s\alpha}} \left(\frac{1}{ \pi  a^2 \gamma }\right)^{\frac{d+1}{2}} \Gamma \left(\frac{d+1}{2}\right) \zeta (d+1).
\end{eqnarray}

\subsection{The Case of $Lv_1$ and $Lv_2$ Lorentz Violations}\label{subsec.ZeroEn.Lv1}
\begin{figure}[th] \hspace{0.2cm}\includegraphics[width=8.5cm]{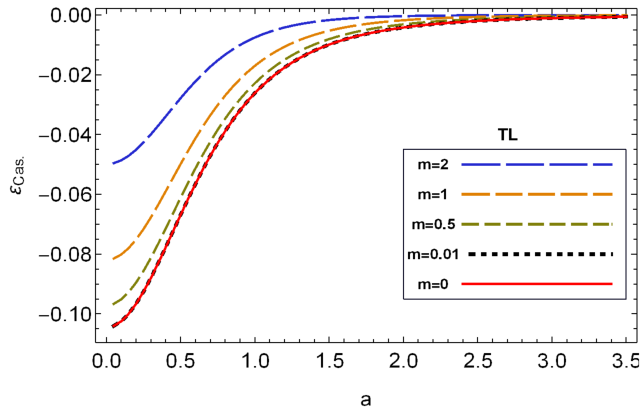}\includegraphics[width=8.5cm]{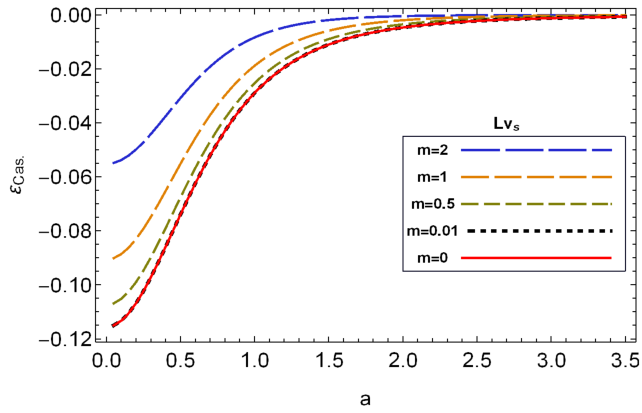}
\hspace{0.2cm}\includegraphics[width=8.5cm]{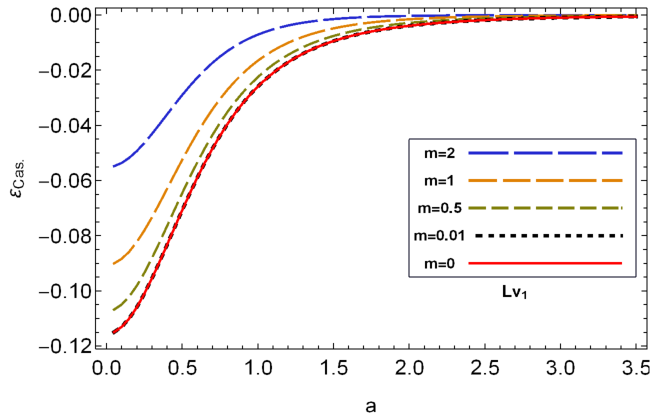}\includegraphics[width=8.5cm]{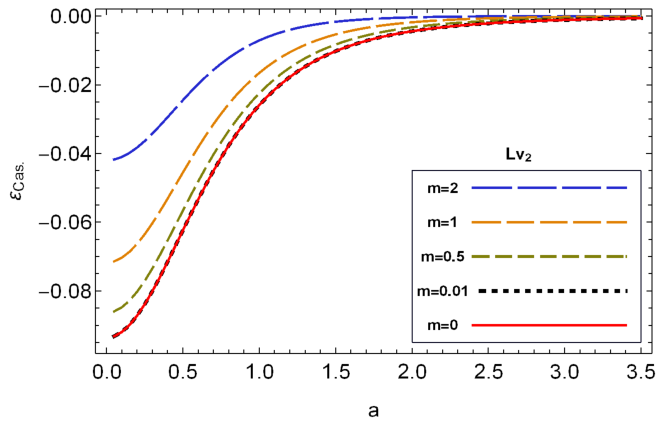}
\caption{\label{PLOT.ZO.MASSIVE.MASSLESS.1} \small
  The leading-order Casimir energy density as a function of $a$ for different mass values in three spatial dimensions, has been plotted. Each figure illustrates a distinct case of Lorentz violation (\emph{e.g.} TL, $Lv_s$, $Lv_1$, and $Lv_2$). It is apparent that the sequence of the massive cases converges swiftly to the massless scenario, with negligible discrepancies observed between the figures of the massive cases for $m\leq0.01$ and the massless case. For all the plots, the parameters were set with $\alpha=0.1$ and $h=1$. }
\end{figure}

To derive the leading-order Casimir energy density for a massive scalar field with $Lv_1$ and $Lv_2$ Lorentz violations under helical boundary condition, we begin with Eqs. (\ref{Wave.number.SLx}) and (\ref{Cas.En.Formal.Form}). Therefore, we obtain
\begin{eqnarray}\label{Zero.Cas.En.5}
      \mathcal{E}^{(0)}_{\mbox{\tiny Cas.}} &=&\frac{1}{2a}\int
      \frac{\prod_{i=2}^{d}dk_i}{(2\pi)^{d-1}}
      \sum_{n=-\infty}^{\infty}\omega_n
      -\frac{1}{2\sqrt{1-\alpha}}\int_{0}^{\infty}
      \frac{\Omega_{d}\kappa^{d-1}d\kappa}{(2\pi)^{d}}
      \big[\kappa^2+m^2\big]^{1/2}.
\end{eqnarray}
The values of the wave-number, written in the aforementioned equation, for the cases $Lv_1$ and $Lv_2$ differ from each other. Whereas for the case of $Lv_1$ Lorentz symmetry breaking, the wave number is $\omega_n=\big[(1-\alpha)k_1^2+k_2^2+...+k_d^2+m^2\big]^{1/2}$, for the case of $Lv_2$, the wave-number is to be substituted with $\omega_n=\big[k_1^2+(1-\alpha)k_2^2+...+k_d^2+m^2\big]^{1/2}$. In the following, upon substituting the wavevector as $k_1=\frac{2n\pi}{a}+k_2 r$ and implementing changing of variables for integrals, Eq. (\ref{Zero.Cas.En.5}) converts to:
\begin{eqnarray}\label{Zero.Cas.En.6}
    \mathcal{E}^{(0)}_{\mbox{\tiny Cas.}} &=&\frac{(1-\alpha)^{d/2}}{2a(2\pi)^{d-1}
       \mathcal{A}_c^{\frac{d+1}{2}}}
       \int\prod_{i=2}^{d}dz_i
      \sum_{n=-\infty}^{\infty}
      \bigg[\Big(\frac{2n\pi}{a}\Big)^2+z_2^2+...+z_d^2+M^2\bigg]^{1/2}\nonumber\\
      &&-\frac{m^{d+1}\Omega_{d}}{2(2\pi)^{d}\sqrt{1-\alpha}}\int_{0}^{\infty}
      \xi^{d-1}\sqrt{\xi^2+1}d\xi,
\end{eqnarray}
where $M=m\sqrt{\frac{\mathcal{A}_c}{1-\alpha}}$ and $\mathcal{A}_c=\gamma-\alpha r^2+c\alpha(r^2-1)$. When $c=0$, we obtain the Casimir energy for the $Lv_1$ case and $c=1$ refers to the case $Lv_2$ of the Lorentz symmetry breaking. To regularize divergences hidden in the summation form of Eq. (\ref{Zero.Cas.En.6}) we utilize the APSF provided in Eq.(\ref{APSF}). The APSF separates this summation expression into two parts. Based on Eq.(\ref{APSF}), the first part is the integral term that is divergent and its contribution would be analytically canceled by last term of Eq. (\ref{Zero.Cas.En.6}). Therefore, there is no contribution for them remain in the Casimir energy. Only the second part of the APSF, which is the branch-cut term, remains. As a result, we obtain:
\begin{eqnarray}\label{branch.cut.LV1}
         \mathcal{E}^{(0)}_{\mbox{\tiny Cas.}}=
         \frac{-2(1-\alpha)^{d/2}\Omega_{d-1}}{(2\pi)^d\mathcal{A}_c^{\frac{d+1}{2}}}\int_{0}^{\infty}z^{d-2}dz
         \int_{\sqrt{z^2+M^2}}^{\infty}\frac{\big[T^2-z^2-M^2\big]^{1/2}}
         {e^{Ta}-1}dT,
\end{eqnarray}
where $z=(z_2,z_3,...,z_d)$. Now, changing the order of integrations, and using the expression $\frac{1}{e^{aT}-1}=\sum_{j=1}^{\infty}e^{-jaT}$ yields the expression for the Casimir energy density in the case of $Lv_1$/$Lv_2$ for Lorentz violating massive scalar field with helix boundary condition. The expression is as follows:
\begin{eqnarray}\label{Zero.Cas.En.7}
     \mathcal{E}^{(0)}_{\mbox{\tiny Cas.}} &=&-2(1-\alpha)^{d/2}\left(\frac{M}{2\pi a\mathcal{A}_c}\right)
     ^{\frac{d+1}{2}}\sum_{j=1}^{\infty}
     \frac{K_{\frac{d+1}{2}}(Maj)}{j^{\frac{d+1}{2}}}.
\end{eqnarray}
In three spatial dimensions, the aforementioned result is consistent with those reported in Ref. \cite{Thermal.Cas.1}. To calculate the Casimir energy for the massless scalar field, one needs to refer back to Eq. (\ref{Zero.Cas.En.5}) and set the mass $m$ to 0. Subsequently, repeating the process outlined in this subsection yields the following expression:
\begin{eqnarray}\label{massless.Zero.Cas.Lv1.Lv2.}
     \mathcal{E}^{(0)}_{\mbox{\tiny Cas.}} \buildrel\mbox{\hspace{0.2cm}$m\to0$\hspace{0.2cm}}\over\longrightarrow-(1-\alpha )^{d/2}\left(\frac{1}{\pi a^2 \mathcal{A}_c}\right)^{\frac{d+1}{2}}  \Gamma \left(\frac{d+1}{2}\right)\zeta(d+1),
\end{eqnarray}
Typically, cross-verifying the Casimir energy results through the language of diagrams enhances confidence in the accuracy of the obtained outcomes. Hence, in Fig. (\ref{PLOT.ZO.MASSIVE.MASSLESS.1}), we have plotted the Casimir energy density as a function of $a$. In this figure, a series of plots corresponding to $m = {1, 0.75, 0.5, 0}$ was presented. These plots illustrate the trend wherein the Casimir energy for massive cases rapidly converges to the massless limit as $m\to 0$. This convergence behavior aligns with previously reported findings and is consistent with physical expectations,\cite{Thermal.Cas.1}.

\section{Thermal Correction}\label{Sec:TC}
In this section, we have computed the thermal correction to the Casimir energy for both massive and massless Lorentz-violating scalar fields with helix boundary condition in $d+1$ dimensions. Because of the distinct directions of Lorentz violation, we have separated the details of the computations into the following subsections. Subsection \ref{subsec.TC.TL&LVs} will address the scenario involving TL/$Lv_s$ Lorentz violation, whereas the cases of $Lv_1$ and $Lv_2$ will be discussed in subsection \ref{subsec.TC.Lv1&Lv2}.

\subsection{The Case of $TL$ and $Lv_s$ Lorentz Violations}\label{subsec.TC.TL&LVs}
Starting with the second parentheses of Eq. (\ref{cas.define.}) and using the thermal correction term of the vacuum energy given in Eq. (\ref{Zero.point&Thermal term.1}), we have
\begin{eqnarray}\label{TC.TL&Lvs.Eq1}
\mathcal{E}^{(T)}_{\mbox{\tiny Cas.}} &=&\frac{1}{\sqrt{1+s\alpha}}\sum_{j=1}^{\infty}\Bigg\{\int
      \frac{\prod_{i=2}^{d}dk_i}{(2\pi)^{d-1}a}
      \sum_{n=-\infty}^{\infty}\big[k_1^2+k_2^2+k_3^2+...+k_d^2+m^2\big]
      ^{1/2}e^{-j\beta\mathcal{R}_s\big[k_1^2+k_2^2+k_3^2+...+k_d^2+m^2\big]
      ^{1/2}}\nonumber\\&&-\int_{0}^{\infty}
      \frac{\Omega_{d}\kappa^{d-1}d\kappa}{(2\pi)^{d}}
      \big[\kappa^2+m^2\big]^{1/2}e^{-j\beta\mathcal{R}_s\big[\kappa^2+m^2\big]^{1/2}} \Bigg\},
\end{eqnarray}
wherein the identity $\frac{1}{e^{\beta(\omega-\mu)}-1}=\sum_{j=1}^{\infty}e^{-j\beta(\omega-\mu)}$ has been employed. We consider the value of the chemical potential for the vacuum energy to be $\mu=0$. Moreover, $\kappa=(\kappa_1,\kappa_2,\kappa_3,\kappa_4,...,\kappa_d)$, $\mathcal{R}_s=\frac{1}{\sqrt{1+\alpha(1+s)/2}}$ and $s=\pm1$. When the parameter $s=1$, we obtain the thermal correction to the Casimir energy for the TL case, while $s=-1$ corresponds to the $Lv_s$ case of Lorentz symmetry breaking. Additionally $\Omega_d=\frac{2\pi^{d/2}}{\Gamma(d/2)}$ represents the spatial angle in $d$ spatial dimensions. By substituting the wavevector $k_1=\frac{2n\pi}{a}+k_2 r$ and performing the changing of variables in integrals, we obtain
\begin{eqnarray}\label{TC.TL&Lvs.Eq2}
       \mathcal{E}^{(T)}_{\mbox{\tiny Cas.}}&=&\frac{1}{(2\pi)^{d-1}a
       \gamma^{\frac{d+1}{2}}\sqrt{1+s\alpha}}\sum_{j=1}^{\infty}
       \int\prod_{i=2}^{d}dz_i
      \sum_{n=-\infty}^{\infty}
      \bigg[\Big(\frac{2n\pi}{a}\Big)^2+z_2^2+...+z_d^2+M^2\bigg]^{1/2}
      \nonumber\\&&\times
      e^{\frac{-j\beta\mathcal{R}_s}{\sqrt{\gamma}}\big[(\frac{2n\pi}{a})^2+z_2^2+z_3^2+...+z_d^2+M^2\big]
      ^{1/2}}-\frac{\Omega_{d}m^{d+1}}{(2\pi)^{d}\sqrt{1+s\alpha}}\sum_{j=1}^{\infty}\int_{0}^{\infty}
      \xi^{d-1}\sqrt{\xi^2+1}e^{-jm\beta\mathcal{R}_s\big[\xi^2+1\big]^{1/2}} d\xi,
\end{eqnarray}
where the parameter $\gamma=1+r^2$ and $M=m\sqrt{\gamma}$. Moreover, changing of variable $\xi=\kappa/m$ was applied. In order to convert the summation over $n$ in Eq. (\ref{TC.TL&Lvs.Eq2}) to integral form, we utilized the APSF as provided in Eq. (\ref{APSF}). Consequently, the following result was obtained:
\begin{eqnarray}\label{TC.TL&Lvs.Eq3}
     \mathcal{E}^{(T)}_{\mbox{\tiny Cas.}} &=&\frac{2}{(2\pi)^{d}
       \gamma^{\frac{d+1}{2}}\sqrt{1+s\alpha}}\sum_{j=1}^{\infty}\Bigg\{
       \int_{0}^{\infty}dX\int_{-\infty}^{\infty}
       \prod_{i=2}^{d}dz_i\big[X^2+z_2^2+...+z_d^2+M^2\big]^{1/2}
       e^{\frac{-j\beta\mathcal{R}_s}{\sqrt{\gamma}}\big[X^2+z_2^2+...+z_d^2+M^2\big]^{1/2}} \nonumber\\&&
      +\mathcal{B}(a)\Bigg\}-\frac{\Omega_{d}m^{d+1}}{(2\pi)^{d}\sqrt{1+s\alpha}}
      \sum_{j=1}^{\infty}\int_{0}^{\infty}
      \xi^{d-1}\sqrt{\xi^2+1}e^{-jm\beta\mathcal{R}_s\big[\xi^2+1\big]^{1/2}}d\xi,
\end{eqnarray}
where $\mathcal{B}(a)$ represent the branch-cut term of the APSF. By choosing $\eta=\frac{1}{M}(X,z_2,z_3,...,z_d)$ in the first expression of Eq. (\ref{TC.TL&Lvs.Eq3}), we obtain:
\begin{eqnarray}\label{TC.TL&Lvs.Eq4}
     \mathcal{E}^{(T)}_{\mbox{\tiny Cas.}} &=&\frac{\Omega_d M^{d+1}}{(2\pi)^{d}
       \gamma^{\frac{d+1}{2}}\sqrt{1+s\alpha}}\sum_{j=1}^{\infty}\Bigg\{
       \int_{0}^{\infty}\eta^{d-1}\big[\eta^2+1\big]^{1/2}
       e^{\frac{-jM\beta\mathcal{R}_s}{\sqrt{\gamma}}\big[\eta^2+1\big]^{1/2}}d\eta \nonumber\\&&
      +\mathcal{B}(a)\Bigg\}-\frac{\Omega_{d}m^{d+1}}{(2\pi)^{d}\sqrt{1+s\alpha}}
      \sum_{j=1}^{\infty}\int_{0}^{\infty}
      \xi^{d-1}\sqrt{\xi^2+1}e^{-jm\beta\mathcal{R}_s\big[\xi^2+1\big]^{1/2}}d\xi.
\end{eqnarray}
The APSF has split the summation over $n$ in Eq. (\ref{TC.TL&Lvs.Eq2}) into two parts. The first part, as reflected in the first term of Eq. (\ref{TC.TL&Lvs.Eq4}), is canceled by the last term arising from the free space. Thus, the only term that remains will be the branch-cut expression, which is as follows:
\begin{eqnarray}\label{TC.TL&Lvs.Eq5}
       \mathcal{E}^{(T)}_{\mbox{\tiny Cas.}} &=&\frac{2i}{(2\pi)^{d-1}a
       \gamma^{\frac{d+1}{2}}\sqrt{1+s\alpha}}\sum_{j=1}^{\infty}
       \int_{0}^{\infty}dt\int_{-\infty}^{\infty}\prod_{\ell=2}^{d}dz_\ell
       \frac{\mathcal{H}(it)-\mathcal{H}(-it)}{e^{2\pi t}-1},
\end{eqnarray}
where
\begin{eqnarray}\label{TC.TL&Lvs.Eq6}
       \mathcal{H}(x)=\Big[\Big(\frac{2\pi x}{a}\Big)^2+z^2_2+...+z^2_d+M^2\Big]^{1/2}
       e^{\frac{-j\beta\mathcal{R}_s}{\sqrt{\gamma}}\big[\big(\frac{2\pi x}{a}\big)^2+z^2_2+...+z^2_d+M^2\big]^{1/2}}.\nonumber
\end{eqnarray}
Upon changing the variable $T=\frac{2\pi t}{a}$, setting $p^2=z^2+M^2$, and denoting $z$ as the vector $z=(z_2,z_3,...,z_d)$ the aforementioned equation is transformed as follows:
\begin{eqnarray}\label{TC.TL&Lvs.Eq7}
       \mathcal{E}^{(T)}_{\mbox{\tiny Cas.}} &=&\frac{-4\Omega_{d-1}}{(2\pi)^{d}
       \gamma^{\frac{d+1}{2}}\sqrt{1+s\alpha}}\sum_{j=1}^{\infty}
       \int_{p}^{\infty}dT\int_{M}^{\infty}dp\frac{p(p^2-M^2)^{\frac{d-3}{2}}
       \sqrt{T^2-p^2}}{e^{aT}-1}\cos\Big(\frac{j\beta\mathcal{R}_s}{\sqrt{\gamma}}\sqrt{T^2-p^2}\Big).
\end{eqnarray}
Now, changing of variable $x^2=(T/p)^2-1$ converts the above equation to:
\begin{eqnarray}\label{TC.TL&Lvs.Eq8}
     \mathcal{E}^{(T)}_{\mbox{\tiny Cas.}} &=&\frac{-4\Omega_{d-1}}{(2\pi)^{d}
       \gamma^{\frac{d+1}{2}}\sqrt{1+s\alpha}}\sum_{j=1}^{\infty}
       \int_{M}^{\infty}dp\int_{0}^{\infty}dx\frac{x^2}{\sqrt{x^2+1}}\frac{p^3(p^2-M^2)^{\frac{d-3}{2}}
       }{e^{ap\sqrt{x^2+1}}-1}\cos\Big(\frac{j\beta\mathcal{R}_s px}{\sqrt{\gamma}}\Big).
\end{eqnarray}
Utilizing expression $1/(e^{ap\sqrt{x^2+1}}-1)=\sum_{n=1}^{\infty}e^{-anp\sqrt{x^2+1}}$ within Eq. (\ref{TC.TL&Lvs.Eq8}) and subsequently performing the integration with respect to $x$ results in
\begin{eqnarray}\label{TC.TL&Lvs.Eq9}
         \mathcal{E}^{(T)}_{\mbox{\tiny Cas.}} &=&\frac{-4\Omega_{d-1}}{(2\pi)^{d}
       \gamma^{\frac{d+1}{2}}\sqrt{1+s\alpha}}\sum_{n=1}^{\infty}
       \sum_{j=1}^{\infty}
       \frac{\partial}{\partial\tilde{\alpha}_j}\left[\int_{M}^{\infty}p^2(p^2-M^2)^{\frac{d-3}{2}}
     \frac{\tilde{\alpha}_j K_1\Big(p\sqrt{\tilde{\alpha}_j^2+(an)^2}\Big)}{\sqrt{\tilde{\alpha}_j^2+(an)^2}}dp\right],
\end{eqnarray}
where $\tilde{\alpha}_j=j\beta\mathcal{R}_s/\sqrt{\gamma}$. In order to compute the remaining integral in Eq. (\ref{TC.TL&Lvs.Eq9}), we changed the variable $\eta=p\sqrt{\tilde{\alpha}_j^2+(an)^2}$. This allowed us to obtain the thermal correction to the Casimir energy corresponding to the case of TL/$Lv_s$ for a Lorentz-violating massive scalar field in $d+1$ dimensions, as follows:
\begin{eqnarray}\label{TC.TL&Lvs.Eq10}
         \mathcal{E}^{(T)}_{\mbox{\tiny Cas.}} &=&\frac{-4}{\sqrt{1+s\alpha}} \left(\frac{M}{2\pi\gamma}\right)^{\frac{d+1}{2}}
       \sum_{n=1}^{\infty}
       \sum_{j=1}^{\infty}\frac{\partial}{\partial\tilde{\alpha}_j}
     \left[\frac{\tilde{\alpha}_j K_{\frac{d+1}{2}}\Big(M\sqrt{\tilde{\alpha}_j^2+(an)^2}\Big)}
     {\big[\tilde{\alpha}_j^2+(an)^2\big]^{\frac{d+1}{4}}}\right].
\end{eqnarray}
Typically, the massless limit of the field serves as a vital benchmark for verifying the accuracy of the solution presented in Eq. (\ref{TC.TL&Lvs.Eq10}). In order to compute the thermal correction to the Casimir energy for a massless scalar field, we return to Eq. (\ref{TC.TL&Lvs.Eq1}) and specify the mass to be zero ($m=0$). By carefully executing the previously detailed calculation steps, we are able to determine the following outcome:
\begin{eqnarray}\label{massless.Thermal.Cas.TL.Lvs.}
    \mathcal{E}^{(T)}_{\mbox{\tiny Cas.}} \buildrel {m\to0}\over\longrightarrow\frac{-2\Gamma(\frac{d+1}{2})}{\sqrt{1+s\alpha}} \left(\frac{1}{\pi\gamma}\right)^{\frac{d+1}{2}}
       \sum_{n=1}^{\infty}
       \sum_{j=1}^{\infty}\frac{\partial}{\partial\tilde{\alpha}_j}
     \left[\frac{\tilde{\alpha}_j}
     {\big[\tilde{\alpha}_j^2+(an)^2\big]^{\frac{d+1}{2}}}\right].
\end{eqnarray}
Figure (\ref{PLOT.THermal.MASSIVE.MASSLESS}) showcases a series of plots depicting the thermal correction to the Casimir energy for both massive and massless Lorentz-violating scalar fields, with varying mass values of $m=\{1,0.75,0.5,0\}$. These plots illustrate that as the mass parameter approaches zero, the Casimir energy for cases with mass converges swiftly to that of the massless limit. This pattern of behavior for the Casimir energy is acceptable on theoretical physical principles\,\cite{Thermal.Cas.1,Thermal.Cas.2}.

\subsection{The Case of $Lv_1$ and $Lv_2$ Lorentz Violations}\label{subsec.TC.Lv1&Lv2}
\begin{figure}[th] \hspace{0.2cm}\includegraphics[width=8.5cm]{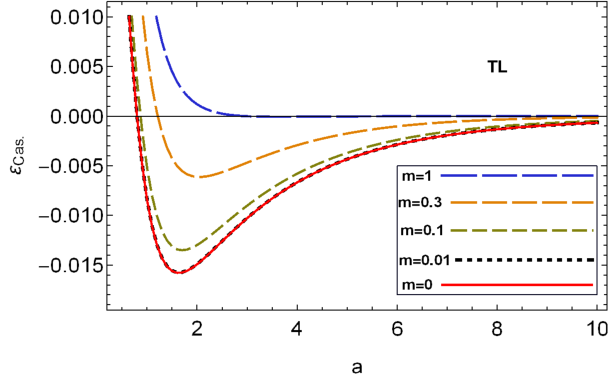}\includegraphics[width=8.5cm]{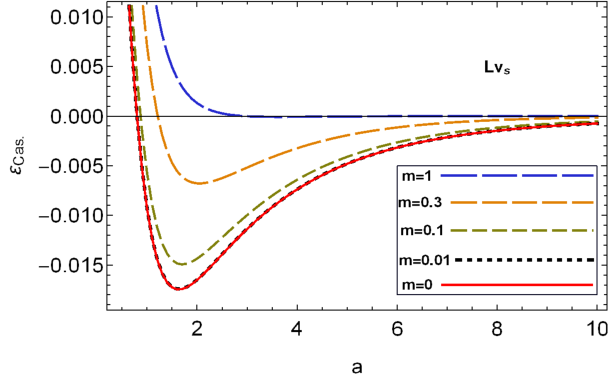}
\hspace{0.2cm}\includegraphics[width=8.5cm]{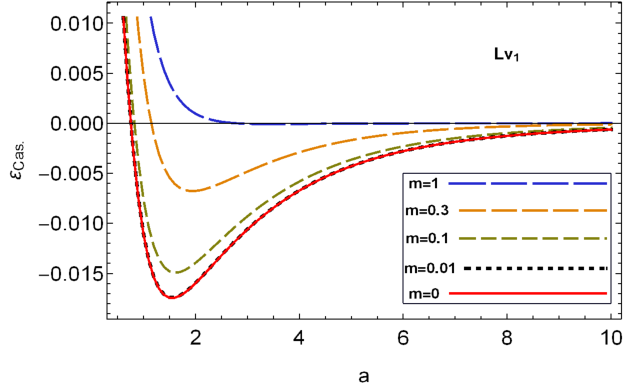}\includegraphics[width=8.5cm]{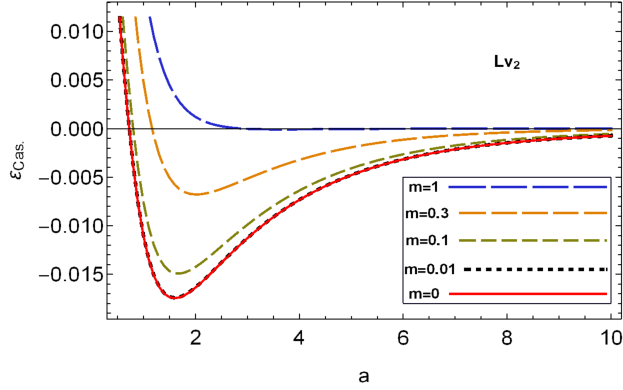}
\caption{\label{PLOT.THermal.MASSIVE.MASSLESS} \small
The thermal correction to the Casimir energy density has been graphically depicted as a function of $a$ in three spatial dimensions, considering various mass values. Each figure corresponds to a distinct case of Lorentz violation, such as TL, $Lv_s$, $Lv_1$, and $Lv_2$. The observed trend reveals a rapid convergence of the sequence of massive cases to the massless case. Notably, there is an inconsequential difference between the figures depicting massive cases for $m\leq0.01$ and the figure representing the massless case. Throughout all plots, the parameters were consistently set as $\mu=0$, $\beta=0.1$, $\alpha=0.1$ and $h=1$.}
\end{figure}

\begin{figure}[th] \hspace{0.2cm}\includegraphics[width=8.5cm]{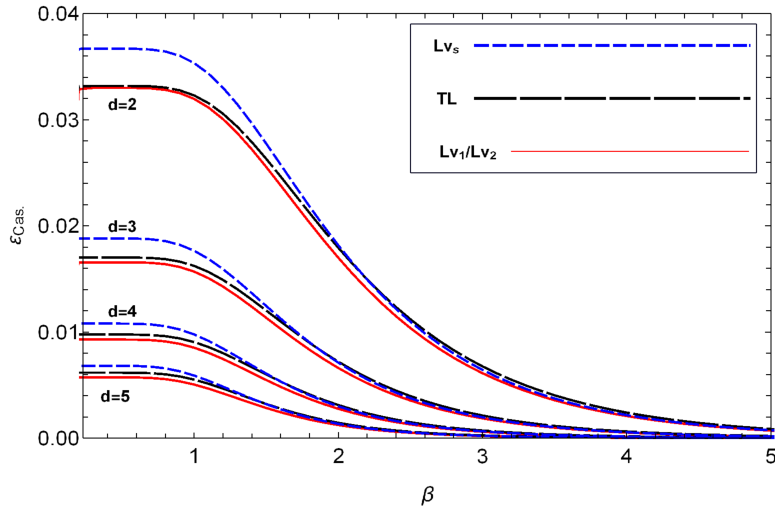}
\caption{\label{PLOT.BETA} \small
The thermal correction to the Casimir energy density has been graphically depicted as a function of $\beta$ for spatial dimensions $d=\{2,3,4,5\}$. Distinct directions of Lorentz symmetry breaking, such as TL, $Lv_s$, $Lv_1$, and $Lv_2$ were distinctly displayed in the plot. The observed trend reveals that by increasing the parameter $\beta$, which corresponds to decreasing the temperature, the thermal correction to the Casimir energy decreases to zero. Throughout all plots, the parameters were consistently set as $\alpha=0.1$, $\mu=0$, $m=1$ and $h=1$.}
\end{figure}

To present the calculation of the thermal correction to the Casimir energy for massive $Lv_1/Lv_2$ Lorentz-violating scalar field, we begin with the expression contained within the second set of parentheses in Eq. (\ref{cas.define.}). Then, employing Eqs. (\ref{Wave.number.SLx}) and (\ref{Zero.point&Thermal term.1}), we proceed as follows:
\begin{eqnarray}\label{TC.Lv1&Lv2.Eq1}
\mathcal{E}^{(T)}_{\mbox{\tiny Cas.}} &=&\frac{1}{(2\pi)^{d-1}a}\sum_{j=1}^{\infty}\int
      \prod_{i=2}^{d}dk_i
      \sum_{n=-\infty}^{\infty}\omega_n  e^{-j\beta\omega_n}-
      \frac{\Omega_{d}}{(2\pi)^{d}\sqrt{1-\alpha}}
      \sum_{j=1}^{\infty}\int_{0}^{\infty}
      \kappa^{d-1}
      \sqrt{\kappa^2+m^2}e^{-j\beta\sqrt{\kappa^2+m^2}}d\kappa,
\end{eqnarray}
where $\mu=0$, $\kappa=(\kappa_1,\kappa_2,\kappa_3,\kappa_4,...,\kappa_d)$, and $\Omega_d=\frac{2\pi^{d/2}}{\Gamma(d/2)}$ is the spatial angle in $d$ spatial dimensions. The wave-number values specified in the earlier equation vary between the cases for $Lv_1$ and $Lv_2$. In the case of $Lv_1$ Lorentz violation, the wave-number is designated as $\omega_n=\big[(1-\alpha)k_1^2+k_2^2+...+k_d^2+m^2\big]^{1/2}$. Whereas, in the context of $Lv_2$ Lorentz violation, the wave-number should be replaced with $\omega_n=\big[k_1^2+(1-\alpha)k_2^2+...+k_d^2+m^2\big]^{1/2}$. Now, utilizing the dispersion relation for each case of Lorentz violation,  and substituting the wavevector as $k_1=\frac{2n\pi}{a}+k_2 r$, coupled with a changing of variables for integrals, the aforementioned equation transforms into:
\begin{eqnarray}\label{TC.Lv1&Lv2.Eq2}
       \mathcal{E}^{(T)}_{\mbox{\tiny Cas.}}&=&\frac{(1-\alpha)^{d/2}}{(2\pi)^{d-1}a
       \mathcal{A}_{c}^{\frac{d+1}{2}}}\sum_{j=1}^{\infty}
       \int\prod_{i=2}^{d}dZ_i
      \sum_{n=-\infty}^{\infty}
      \bigg[\Big(\frac{2n\pi}{a}\Big)^2+Z_2^2+Z_3^2+...+Z_d^2+M^2\bigg]^{1/2}
      \nonumber\\&&\times
      e^{-j\beta\sqrt{\frac{1-\alpha}{\mathcal{A}_c}}
      \big[(\frac{2n\pi}{a})^2+Z_2^2+Z_3^2+...+Z_d^2+M^2\big]
      ^{1/2}}-\frac{\Omega_{d}m^{d+1}}{(2\pi)^{d}\sqrt{1-\alpha}}
      \sum_{j=1}^{\infty}\int_{0}^{\infty}
      \xi^{d-1}\sqrt{\xi^2+1}e^{-jm\beta\sqrt{\xi^2+1}} d\xi,\nonumber\\
\end{eqnarray}
where the changing of variables $\xi=\kappa/m$, $Z_2=\frac{\mathcal{A}_c}{\sqrt{1-\alpha}}\left[k_2+\frac{2n\pi r(1-\alpha)}{a\mathcal{A}_c}\right]$ and $(Z_3,...,Z_d)=\sqrt{\frac{\mathcal{A}_c}{1-\alpha }}\left(k_3,...,k_d\right)$ were applied. Additionaly, the parameters $\mathcal{A}_c=\gamma-\alpha r^2+c\alpha(r^2-1)$, and the parameter $M$ was introduced as $M=m\sqrt{\frac{\mathcal{A}_c}{1-\alpha}}$. For the index $c=0$, the thermal correction to the Casimir energy for the case of $Lv_1$ would be obtained and $c=1$ refers to the case of $Lv_2$ Lorentz violation. In order to convert the summation over $n$ into integral form in Eq. (\ref{TC.Lv1&Lv2.Eq2}), we utilized the APSF described in Eq. (\ref{APSF}). Therefore, we obtained:
\begin{eqnarray}\label{TC.Lv1&Lv2.Eq3}
     \mathcal{E}^{(T)}_{\mbox{\tiny Cas.}} &=&\frac{2(1-\alpha)^{d/2}}{(2\pi)^{d}
       \mathcal{A}_c^{\frac{d+1}{2}}}\sum_{j=1}^{\infty}
       \Bigg\{
       \int_{0}^{\infty}dX\int_{-\infty}^{\infty}
       \prod_{i=2}^{d}dZ_i\big[X^2+Z_2^2+...+Z_d^2+M^2\big]^{1/2}
       e^{-j\beta\sqrt{\frac{1-\alpha}{\mathcal{A}_c}}\big[X^2+Z_2^2+...+Z_d^2+M^2\big]^{1/2}} \nonumber\\&&
      +\mathcal{B}(a)\Bigg\}-\frac{\Omega_{d}m^{d+1}}{(2\pi)^{d}\sqrt{1-\alpha}}
      \sum_{j=1}^{\infty}\int_{0}^{\infty}
      \xi^{d-1}\sqrt{\xi^2+1}e^{-jm\beta\sqrt{\xi^2+1}}d\xi,
\end{eqnarray}
where $\mathcal{B}(a)$ denotes the branch-cut term of APSF. Now, changing of variable $\eta=\frac{1}{M}(X,Z_2,Z_3,...,Z_d)$ in the initial expression of Eq. (\ref{TC.Lv1&Lv2.Eq3}) results the following form
\begin{eqnarray}\label{TC.Lv1&Lv2.Eq4}
     \mathcal{E}^{(T)}_{\mbox{\tiny Cas.}} &=&\frac{(1-\alpha)^{d/2} M^{d+1}\Omega_d}{(2\pi)^{d}
       \mathcal{A}_c^{\frac{d+1}{2}}}\sum_{j=1}^{\infty}\Bigg\{
       \int_{0}^{\infty}\eta^{d-1}\sqrt{\eta^2+1}
       e^{-jM\beta\sqrt{\frac{1-\alpha}{\mathcal{A}_c}}\sqrt{\eta^2+1}}d\eta \nonumber\\&&
      +\mathcal{B}(a)\Bigg\}-\frac{\Omega_{d}m^{d+1}}{(2\pi)^{d}\sqrt{1-\alpha}}
      \sum_{j=1}^{\infty}\int_{0}^{\infty}
      \xi^{d-1}\sqrt{\xi^2+1}e^{-jm\beta\sqrt{\xi^2+1}}d\xi.
\end{eqnarray}
The first part of Eq. (\ref{TC.Lv1&Lv2.Eq4}) is analytically canceled out by the last term arising from the free space. Consequently, the only remaining term pertains to the branch-cut term $\mathcal{B}(a)$, and we have:
\begin{eqnarray}\label{TC.Lv1&Lv2.Eq5}
       \mathcal{E}^{(T)}_{\mbox{\tiny Cas.}} &=&\frac{2i(1-\alpha)^{d/2}}{(2\pi)^{d-1}
       \mathcal{A}_c^{\frac{d+1}{2}}a}\sum_{j=1}^{\infty}
       \int_{0}^{\infty}dt\int_{-\infty}^{\infty}\prod_{\ell=2}^{d}dZ_\ell
       \frac{\mathcal{H}(it)-\mathcal{H}(-it)}{e^{2\pi t}-1},
\end{eqnarray}
where
\begin{eqnarray}\label{TC.Lv1&Lv2.Eq6}
       \mathcal{H}(x)=\Big[\Big(\frac{2\pi x}{a}\Big)^2+Z^2_2+...+Z^2_d+M^2\Big]^{1/2}
       e^{-j\beta\sqrt{\frac{1-\alpha}{\mathcal{A}_c}}
       \big[\big(\frac{2\pi x}{a}\big)^2+Z^2_2+...+Z^2_d+M^2\big]^{1/2}}.\nonumber
\end{eqnarray}
Changing of variables $T=\frac{2\pi t}{a}$, $p^2=Z^2+M^2$, and $Z=(Z_2,Z_3,...,Z_d)$ convert the above equation to:
\begin{eqnarray}\label{TC.Lv1&Lv2.Eq7}
       \mathcal{E}^{(T)}_{\mbox{\tiny Cas.}} &=&\frac{-4\Omega_{d-1}(1-\alpha)^{d/2}}{(2\pi)^{d}
       \mathcal{A}_c^{\frac{d+1}{2}}}
       \sum_{j=1}^{\infty}
       \int_{p}^{\infty}dT\int_{M}^{\infty}
       dp\frac{p(p^2-M^2)^{\frac{d-3}{2}}
       \sqrt{T^2-p^2}}{e^{aT}-1}
       \cos\Big(j\beta\sqrt{\frac{1-\alpha}{\mathcal{A}_c}}\sqrt{T^2-p^2}\Big)
\end{eqnarray}
Upon other change of the variable to $x^2=(T/p)^2-1$ and subsequent integration over $x$, the resulting expression is as follows:
\begin{eqnarray}\label{TC.Lv1&Lv2.Eq9}
         \mathcal{E}^{(T)}_{\mbox{\tiny Cas.}} &=&\frac{-4\Omega_{d-1}(1-\alpha)^{d/2}}{(2\pi)^{d}
       \mathcal{A}_c^{\frac{d+1}{2}}}\sum_{n=1}^{\infty}
       \sum_{j=1}^{\infty}
       \frac{\partial}{\partial\tilde{\alpha}_j}\left[\int_{M}^{\infty}p^2(p^2-M^2)^{\frac{d-3}{2}}
     \frac{\tilde{\alpha}_j K_1\Big(p\sqrt{\tilde{\alpha}_j^2+(an)^2}\Big)}{\sqrt{\tilde{\alpha}_j^2+(an)^2}}dp\right],
\end{eqnarray}
where $\tilde{\alpha}_j=j\beta\sqrt{\frac{1-\alpha}{\mathcal{A}_c}}$. The remaining integral in Eq. (\ref{TC.Lv1&Lv2.Eq9}) after the change of variable $\eta=p\sqrt{\tilde{\alpha}_j^2+(an)^2}$ is performed. The ultimate result yields the thermal correction to the Casimir energy density for $Lv_1/Lv_2$ Lorentz-violating massive scalar field in $d+1$ dimensions, as expressed below:
\begin{eqnarray}\label{TC.Lv1&Lv2.Eq10}
\mathcal{E}^{(T)}_{\mbox{\tiny Cas.}} &=&-4(1-\alpha)^{d/2} \left(\frac{M}{2\pi\mathcal{A}_c}\right)^{\frac{d+1}{2}}
       \sum_{n=1}^{\infty}
       \sum_{j=1}^{\infty}\frac{\partial}{\partial\tilde{\alpha}_j}
     \left[\frac{\tilde{\alpha}_j K_{\frac{d+1}{2}}\Big(M\sqrt{\tilde{\alpha}_j^2+(an)^2}\Big)}
     {\big[\tilde{\alpha}_j^2+(an)^2\big]^{\frac{d+1}{4}}}\right].
\end{eqnarray}
Reiterating the computation process starting from Eq. (\ref{TC.Lv1&Lv2.Eq1}) and considering the field mass as $m=0$ leads to the derivation of the massless limit of the thermal correction to the Casimir energy for $Lv_1$/$Lv_2$ Lorentz-violating scalar field. The resulting expression is as follows:
\begin{eqnarray}\label{massless.Thermal.Cas.Lv1.Lv2.}
    \mathcal{E}^{(T)}_{\mbox{\tiny Cas.}} \buildrel {m\to0}\over\longrightarrow\frac{-2\Gamma(\frac{d+1}{2})}{(1-\alpha)^{-d/2}} \left(\frac{1}{\pi\mathcal{A}_c}\right)^{\frac{d+1}{2}}
       \sum_{n=1}^{\infty}
       \sum_{j=1}^{\infty}\frac{\partial}{\partial\tilde{\alpha}_j}
     \left[\frac{\tilde{\alpha}_j}
     {\big[\tilde{\alpha}_j^2+(an)^2\big]^{\frac{d+1}{2}}}\right].
\end{eqnarray}
Fig. (\ref{PLOT.THermal.MASSIVE.MASSLESS}) serves as a clear demonstration of the consistency between the outcomes derived in Eqs. (\ref{TC.Lv1&Lv2.Eq10}) and (\ref{massless.Thermal.Cas.Lv1.Lv2.}). Within this figure, the thermal correction to the Casimir energy for both massive and massless scalar fields has been graphically represented. The convergence of the plotted curves for the massive cases, as the parameter $m$ approaches zero, toward the plot corresponding to the massless case is in accordance with established physical principles. Another crucial observation from the displayed plots in Fig. (\ref{PLOT.THermal.MASSIVE.MASSLESS}) is the variation in the sign of the thermal correction term in the Casimir energy. For all directions of Lorentz violation, by decreasing the mass value, this sign change occurs at lower values of $a$. In Fig. (\ref{PLOT.BETA}), the thermal correction to the Casimir energy density for dimensions $d=\{2,3,4,5\}$ as a function of $\beta$ is presented. The plot reveals that as the temperature decreases, the Casimir energy density also decreases. Furthermore, in lower dimensions, this downward trend exhibits a steeper slope.

\section{Radiative Correction}\label{Sec:RC}
In this section, we explore the radiative correction to the Casimir energy for a Lorentz-violating scalar field, both massive and massless, within the context of the $\phi^4$ theory in $d$ spatial dimensions. To achieve this, two analogous systems with helical boundary conditions were examined. For the original system, referred to as ``System A,'' the expression provided in Eq. (\ref{Helix.BC.}) is utilized to emulate the helical structure. In contrast, for the second system, denoted as ``System B,'' the following condition is upheld:
\begin{eqnarray}\label{Helix.BC.2}
\phi(t,x_1+b,x_2,...,x_d)=\phi(t,x_1,x_2+h,...,x_d),
\end{eqnarray}
Now, to obtain the third parenthesis of Eq. (\ref{cas.define.}), we define the radiative correction to the Casimir energy density as follows:
\begin{eqnarray}\label{RC.Cas.def}
\mathcal{E}_{\mbox {\tiny Cas.}}^{(R)}=\lim_{b\to\infty}\Big[\mathcal{E}^{(R)}_{\mbox {\tiny Vac.}}(a)-\mathcal{E}^{(R)}_{\mbox {\tiny Vac.}}(b)\Big].
\end{eqnarray}
Here, $\mathcal{E}^{(R)}_{\mbox {\tiny Vac.}}(a)$ represents the first-order vacuum energy ($\mathcal{O}(\lambda)$) corresponding to the System A. Additionally, $\mathcal{E}^{(R)}_{\mbox {\tiny Vac.}}(b)$ pertains to the vacuum energy of the system B.
We maintain that this definition of Casimir energy is based on conventional definitions, differing in that the vacuum energy of Minkowski space was substituted by that of System B. We assert that as $b\to\infty$, the vacuum energy characteristics of System B converge to those of Minkowski space. Consequently, employing System B does not distort the standard definition of Casimir energy but only adds an additional parameter in the calculation process. These extra parameters serve as regulators, aiding in a more transparent elimination of infinities. Considering the different directions of Lorentz-violation, akin to the preceding sections, details of computations have been compartmentalized into the subsequent subsections. Subsection \ref{subsec.RC.TL&LVs} will delve into cases involving TL and $Lv_s$ Lorentz violation, while subsection \ref{subsec.RC.Lv1&Lv2} will focus on the cases of $Lv_1$ and $Lv_2$.
\subsection{The Case of TL and $Lv_s$ Lorentz Violations}\label{subsec.RC.TL&LVs}
Starting with Eqs. (\ref{First.Order.Vacuum.En.}) and (\ref{Greens.Function.TLHBC}) for the case of TL/$Lv_s$ Lorentz-violating, after performing the integration over the wave-number $\omega$, the subtraction of the vacuum energy densities within the brackets of Eq. (\ref{RC.Cas.def}) can be expressed as follows:
\begin{eqnarray}\label{RC.Cas.Eq.1}
     \mathcal{E}_{\mbox {\tiny Vac.}}^{(R)}(a)- \mathcal{E}_{\mbox {\tiny Vac.}}^{(R)}(b)=\frac{-\lambda\pi^2 }{8a^2(2\pi)^{2d}(1+s\alpha)}\Bigg[\sum_{n=-\infty}^{\infty}\int \prod_{i=2}^{d}dk_i
      \Big[k_1^2+k_2^2+k_3^2+...+k_d^2+m^2\Big]^{-1/2}
       \Bigg]^2-\{a\to b\},
\end{eqnarray}
where $s=1$ corresponds to the TL Lorentz violation, and $s=-1$ pertains to the case of $Lv_s$ Lorentz violation. Substitution of the wavevector $k_1=\frac{2n\pi+k_2h}{a}$ and the subsequent change of variables transform the above expression into:
\begin{eqnarray}\label{RC.Cas.Eq.2}
     \mathcal{E}_{\mbox {\tiny Vac.}}^{(R)}(a)- \mathcal{E}_{\mbox {\tiny Vac.}}^{(R)}(b)=\frac{-\lambda\pi^2\gamma_a^{1-d} }{8a^2(2\pi)^{2d}(1+c\alpha)}\Bigg[\sum_{n=-\infty}^{\infty}\int \prod_{i=2}^{d}dz_i
      \Big[\Big(\frac{2n\pi}{a}\Big)^2+z_2^2+z_3^2+...+z_d^2+M^2\Big]^{-1/2}
       \Bigg]^2-\{a\to b\},
\end{eqnarray}
where $\gamma_a=1+(h/a)^2$ and $M=m\sqrt{\gamma_a}$. The summation over $n$ in aforementioned equation renders it divergent. To regularize it, we first apply the APSF as given in Eq. (\ref{APSF}), dividing the summation into two integral forms. This process separates the divergent portion from the convergent contribution, thereby creating a situation in which the infinities can be eliminated with clarity. Consequently, we obtain:
\begin{eqnarray}\label{RC.Cas.Eq.3}
     \mathcal{E}_{\mbox {\tiny Vac.}}^{(R)}(a)- \mathcal{E}_{\mbox {\tiny Vac.}}^{(R)}(b)&=&\frac{-\lambda\pi^2\gamma_{a}^{1-d} }{8a^2(2\pi)^{2d}(1+s\alpha)}\Bigg[\left(\frac{a\Omega_d M^{d-1}}{2\pi}\right)^2\underbrace{ \left(\int_{0}^{\infty}\eta^{d-1}d\eta
      \Big[\eta^2+1\Big]^{-1/2}\right)^2}_{\mathcal{I}(\infty)^2}       +2\left(\frac{a\Omega_d M^{d-1}}{2\pi}\right)\mathcal{I}(\infty)\mathcal{B}(a)
      \nonumber\\&&\hspace{3cm}+\mathcal{B}(a)^2\Bigg]
      -\{a\to b\},
\end{eqnarray}
where $\eta=\frac{1}{M}(\frac{2n\pi}{a},z_2,...,z_d)$. As mentioned earlier, the integral term of APSF ($\mathcal{I}(\infty)$) is divergent and it results the first two terms within the bracket of Eq. (\ref{RC.Cas.Eq.3}) also being divergent. The first term in the bracket of Eq. (\ref{RC.Cas.Eq.3}) would be canceled analytically by its corresponding term pertaining to the second system (System B). However, removing infinities caused by the second term of Eq. (\ref{RC.Cas.Eq.3}) requires a regularization technique. To this end, we have set the upper limit of the integral $\mathcal{I}$ to a specific cutoff value, denoted as $\Lambda_a$. Simultaneously, the upper limit of the integral $\mathcal{I}$ in the second term, pertaining to System B, was set to $\Lambda_b$. We assert that by appropriately adjusting the values for the cutoffs, one can eliminate the divergences originating from the second term in the bracket of the above equation for both systems (System A and B). In accordance with this, the cutoffs $\Lambda_a$ and $\Lambda_b$ were adjusted as follows:
\begin{eqnarray}\label{adjust.cutoff.1}
     \frac{\mathcal{I}(\Lambda_a)}{\mathcal{I}(\Lambda_b)}=
     \frac{a\gamma_a^{\frac{d-1}{2}}\mathcal{B}(b)}
     {b\gamma_b^{\frac{d-1}{2}}\mathcal{B}(a)}.
\end{eqnarray}
By applying this adjustment to the cutoffs, all divergent parts arising from the second term within the bracket of Eq. (\ref{RC.Cas.Eq.3}) would be eliminated. Therefore, Eq. (\ref{RC.Cas.Eq.3}) converts to:
\begin{eqnarray}\label{RC.Cas.Eq.4}
   \mathcal{E}_{\mbox {\tiny vac.}}^{(R)}(a)- \mathcal{E}_{\mbox {\tiny vac.}}^{(R)}(b)&=&\frac{-\lambda\pi^2\gamma_{a}^{1-d} }{8a^2(2\pi)^{2d}(1+c\alpha)}\mathcal{B}(a)^2-\{a\to b\},
\end{eqnarray}
where $\mathcal{B}(a)$ is
\begin{eqnarray}\label{RC.Branch.cut.1}
     \mathcal{B}(a)=\frac{2a\Omega_{d-1}}{\pi}\int_{M}^{\infty}dT
     \int_{0}^{\sqrt{T^2-M^2}}z^{d-2}
     dz\frac{(T^2-z^2-M^2)^{-1/2}}{e^{2Ta}-1},\hspace{1.5cm} z=(z_2,z_3,...,z_d).
\end{eqnarray}
Upon computing the integration in above equation and incorporating the outcomes into Eqs. (\ref{RC.Cas.def},\ref{RC.Cas.Eq.4}), we then evaluate the limit $b\to\infty$. This process leads to the following result,
\begin{eqnarray}\label{RC.Cas.final.TL&Lvs}
         \mathcal{E}^{(R)}_{\mbox{\tiny Cas.}}=\frac{-\lambda\gamma_{a}^{1-d} }{2(2\pi)^{d+1}(1+s\alpha)}
         \left[\sum_{j=1}^{\infty}\Big(\frac{M}{aj}\Big)^{\frac{d-1}{2}}
         K_{\frac{d-1}{2}}(Maj)\right]^2.
\end{eqnarray}
This outcome reveals the first-order radiative correction to the Casimir energy density for the Lorentz-violating massive scalar field, specifically pertaining to both the TL and $Lv_s$ cases. For the massless scalar field, the limit $m\to0$ has been taken. This limiting process transforms the aforementioned outcome into the following result:
\begin{eqnarray}\label{RC.massless.Cas.TL&Lvs}
         \mathcal{E}^{(R)}_{\mbox{\tiny Cas.}}\buildrel {m\to0}\over\longrightarrow\frac{-\lambda\gamma_{a}^{1-d} }{2(2\pi)^{d+1}(1+s\alpha)}\left(\sin^2(\pi d/2)+\frac{\pi}{2}\cos^2(\pi d/2)\right)
         \left[\sum_{j=1}^{\infty}\frac{(d-3)!!}{(aj)^{d-1}}\right]^2.
\end{eqnarray}
\subsection{The Case of $Lv_1$ and $Lv_2$ Lorentz Violations}\label{subsec.RC.Lv1&Lv2}
\begin{figure}[th] \hspace{0.2cm}\includegraphics[width=8.5cm]{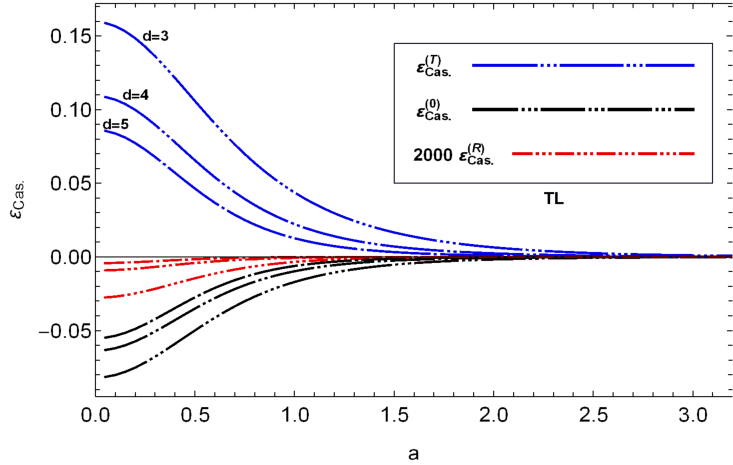}\includegraphics[width=8.5cm]{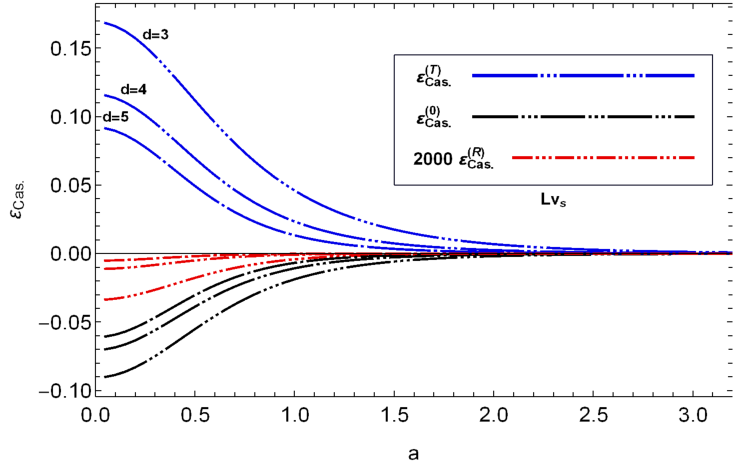}
\hspace{0.2cm}\includegraphics[width=8.5cm]{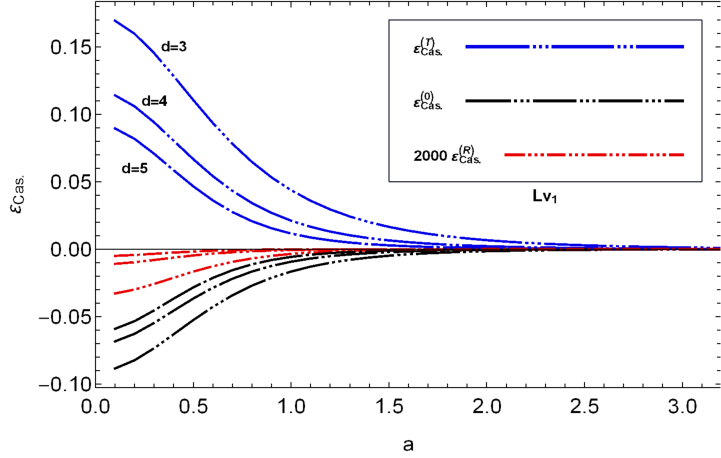}\includegraphics[width=8.5cm]{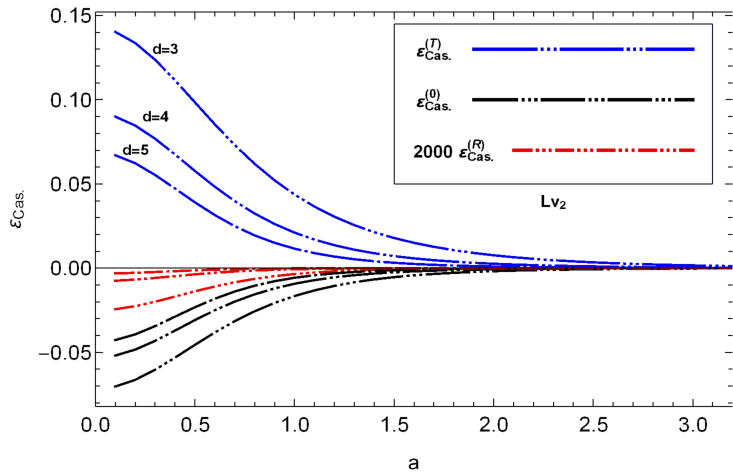}
\caption{\label{3PLOTs} \small
  The figure presents the leading-order Casimir energy density and its thermal and radiative corrections plotted as functions of $a$ for a Lorentz-violated massive scalar field within $d=\{3,4,5\}$ spatial dimensions. The parameters for all plots were fixed at $\alpha=0.1$, $\beta=0.1$, $\mu=0.1$ and $m=1$, and $h=1$.}
\end{figure}

To obtain the radiative correction to the Casimir energy for Lorentz-violating massive scalar field in cases $Lv_1$ and $Lv_2$ we start by,
\begin{eqnarray}\label{RC.Cas.En.Lv1&2.Eq.1}
   \mathcal{E}_{\mbox {\tiny Vac.}}^{(R)}(a)- \mathcal{E}_{\mbox {\tiny Vac.}}^{(R)}(b)=\frac{-\lambda\pi^2(1-\alpha)^{d-2} }{8a^2(2\pi)^{2d}\mathcal{A}_c(a)^{d-1}}\Bigg[\sum_{n=-\infty}^{\infty}\int \prod_{i=2}^{d}dk_i
      \Big[\left(\frac{2n\pi}{a}\right)^2+k_2^2+k_3^2+...+k_d^2+M^2\Big]^{-1/2}
       \Bigg]^2-\{a\to b\},
\end{eqnarray}
where $M=m\sqrt{\frac{\mathcal{A}_c(a)}{1-\alpha}}$ and the parameter $\mathcal{A}_c(a)=\gamma_a-\alpha (h/a)^2+c\alpha((h/a)^2-1)$. Furthermore, the parameter $\gamma_a=1+(h/a)^2$. The parameter $c$ can evalutae two numbers $0$ and $1$. When the parameter is $c=0$ the Lorentz symmetry breaking in the case $Lv_1$ occures and $c=1$ refers to the case of $Lv_2$ of Lorentz violation. To regularize the summation over $n$ in Eq. (\ref{RC.Cas.En.Lv1&2.Eq.1}), the APSF, as given in Eq. (\ref{APSF}), was used. Thus, we obtain:
\begin{eqnarray}\label{RC.Cas.En.Lv1&2.Eq.2}
     \mathcal{E}_{\mbox {\tiny vac.}}^{(R)}(a)- \mathcal{E}_{\mbox {\tiny vac.}}^{(R)}(b)&=&\frac{-\lambda\pi^2(1-\alpha)^{d-2} }{8a^2(2\pi)^{2d}\mathcal{A}_c(a)^{d-1}}\Bigg[\left(\frac{a\Omega_d M^{d-1}}{2\pi}\right)^2\underbrace{ \left(\int_{0}^{\infty}\eta^{d-1}d\eta
      \Big[\eta^2+1\Big]^{-1/2}\right)^2}_{\mathcal{J}(\infty)^2}       +2\left(\frac{a\Omega_d M^{d-1}}{2\pi}\right)\mathcal{J}(\infty)\mathcal{B}(a)
      \nonumber\\&&\hspace{3cm}+\mathcal{B}(a)^2\Bigg]
      -\{a\to b\}.
\end{eqnarray}
The first term in the bracket of Eq. (\ref{RC.Cas.En.Lv1&2.Eq.2}) would be canceled automatically by its corresponding term pertaining to the second system\,(System B). However, removing infinities caused by the second term of Eq. (\ref{RC.Cas.En.Lv1&2.Eq.2}) requires a regularization technique. To this end, we employed the cutoff regularization technique and adjusted the upper limit of the integral $\mathcal{J}$ to a specific cutoff value, denoted as $\Lambda'_a$. Simultaneously, the upper limit of the integral in the second term—pertaining to System B—was set to $\Lambda'_b$. By properly adjusting the cutoffs, all divergences originating from the second term within the brackets of Eq. (\ref{RC.Cas.En.Lv1&2.Eq.2})—pertaining to Systems A and B—will cancel each other out. According to this, we determine the cutoffs by the following relations:
\begin{eqnarray}\label{adjust.cutoff.2}
     \frac{\mathcal{J}(\Lambda'_a)}{\mathcal{J}(\Lambda'_b)}=
     \frac{a\mathcal{B}(b)}
     {b\mathcal{B}(a)}\left[\frac{\mathcal{A}_c(a)}
     {\mathcal{A}_c(b)}\right]^{\frac{d-1}{2}}.
\end{eqnarray}
Therefore, Eq. (\ref{RC.Cas.En.Lv1&2.Eq.2}) is transformed as follows:
\begin{eqnarray}\label{RC.Cas.En.Lv1&2.Eq.3}
  \mathcal{E}_{\mbox {\tiny Vac.}}^{(R)}(a)- \mathcal{E}_{\mbox {\tiny Vac.}}^{(R)}(b)&=&\frac{-\lambda\pi^2(1-\alpha)^{d-2} }{8a^2(2\pi)^{2d}\mathcal{A}_c(a)^{d-1}}\mathcal{B}(a)^2-\{a\to b\}.
\end{eqnarray}
Considering the expression for the branch-cut term provided in Eq. (\ref{RC.Branch.cut.1}) and substituting it in Eq. (\ref{RC.Cas.En.Lv1&2.Eq.3}), the limit $b\to\infty$ leads to the determination of the radiative correction to the Casimir energy density, which can be articulated as follows:
\begin{eqnarray}\label{RC.Cas.final.Lv1&Lv2}
     \mathcal{E}^{(R)}_{\mbox{\tiny Cas.}}=
     \frac{-\lambda(1-\alpha)^{d-2} }{2(2\pi)^{d+1}\mathcal{A}_c(a)^{d-1}}\left[\sum_{j=1}^{\infty}\Big(\frac{M}{aj}\Big)^{\frac{d-1}{2}}
     K_{\frac{d-1}{2}}(Maj)\right]^2.
\end{eqnarray}
Upon careful examination of the expressions represented in Eqs. (\ref{RC.Cas.final.Lv1&Lv2}) and (\ref{Zero.Cas.En.7}), one can establish a pertinent relation between the two. This relationship equation is:
\begin{eqnarray}\label{ERTEBAT.ZO.RC.Lv1&2}
      \mathcal{E}^{(R)}_{\mbox{\tiny Cas.}}(d+2,a)=\frac{-\lambda}{32\pi^2}\mathcal{E}^{(0)}_{\mbox{\tiny Cas.}}(d,a)^2,
\end{eqnarray}
wherein $\mathcal{E}^{(0)}_{\mbox{\tiny Cas.}}(d,a)$ represents the leading-order Casimir energy within $d$ spatial dimensions, while $\mathcal{E}^{(R)}_{\mbox{\tiny Cas.}}(d+2,a)$ refers to the radiative correction to the Casimir energy in $d+2$ spatial dimensions. It is noteworthy that the derivation of above relationship involved the use of two specific equations, (\ref{RC.Cas.final.Lv1&Lv2}) and (\ref{Zero.Cas.En.7}), which pertain to Lorentz symmetry breaking in the $Lv_1$ and $Lv_2$ cases. However, this relationship can be generalized to other cases of Lorentz symmetry breaking, notably the TL and $Lv_s$. Further exploration reveals that it is feasible to deduce a relation between the leading-order Casimir energy, which preserves Lorentz symmetry (where $\alpha=0$), and its radiative correction, which violates Lorentz symmetry (where $\alpha\neq0$). Implementing Eqs. (\ref{Zero.Cas.En.7},\ref{Zero.Cas.En.4}) with the condition of $\alpha=0$ and substituting the result into Eqs. (\ref{RC.Cas.final.TL&Lvs},\ref{RC.Cas.final.Lv1&Lv2}), elucidates their relationship. This relationship expression is:
\begin{eqnarray}\label{ERTEBAT.ZO.RC.Lv1&2.2}
     \mathcal{E}^{(R)}_{\mbox{\tiny Cas.}}(d+2,a)
=\left\{
         \begin{array}{ll}
          \frac{-\lambda}{32\pi^2(1-\alpha)}
     \tilde{\mathcal{E}}^{(0)}_{\mbox{\tiny Cas.}}(d,\tilde{a})^2, & \hbox{where $\tilde{a}=a\sqrt{\frac{\gamma-\alpha r^2}{\gamma(1-\alpha)}}$ for the case of $Lv_1$;} \\
          \frac{-\lambda}{32\pi^2(1-\alpha)}
     \tilde{\mathcal{E}}^{(0)}_{\mbox{\tiny Cas.}}(d,\tilde{a})^2, & \hbox{where $\tilde{a}=a\sqrt{\frac{\gamma-\alpha}{\gamma(1-\alpha)}}$ for the case of $Lv_2$;}\\
 \frac{-\lambda}{32\pi^2(1+\alpha)}
     \tilde{\mathcal{E}}^{(0)}_{\mbox{\tiny Cas.}}(d,a)^2,& \hbox{for the case of TL;}\\
 \frac{-\lambda}{32\pi^2(1-\alpha)}
     \tilde{\mathcal{E}}^{(0)}_{\mbox{\tiny Cas.}}(d,a)^2,& \hbox{for the case of $Lv_s$,}
         \end{array}
  \right.
\end{eqnarray}
where $\tilde{\mathcal{E}}^{(0)}_{\mbox{\tiny Cas.}}(d,a)$ is the leading-order Casimir energy without any Lorentz violation($\alpha=0$). To derive the radiative correction to the Casimir energy in the massless limit of the field, we revisit Eq. (\ref{RC.Cas.En.Lv1&2.Eq.1}) and assign a value of zero to the mass parameter $m$. Iterating the computational process delineated in this subsection, we arrive at the subsequent expression for the radiative correction to the Casimir energy density pertinent to a massless $Lv_1/Lv_2$ Lorentz-violating scalar field:
\begin{eqnarray}\label{RC.massless.Cas.Lv1&2}
     \mathcal{E}^{(R)}_{\mbox{\tiny Cas.}}\buildrel {m\to0}\over\longrightarrow\frac{-\lambda(1-\alpha)^{d-2} }{2(2\pi)^{d+1}\mathcal{A}_c(a)^{d-1}}\left(\sin^2(\pi d/2)+\frac{\pi}{2}\cos^2(\pi d/2)\right)
         \left[\sum_{j=1}^{\infty}\frac{(d-3)!!}{(aj)^{d-1}}\right]^2.
\end{eqnarray}
\begin{figure}[th] \hspace{0.2cm}\includegraphics[width=8.5cm]{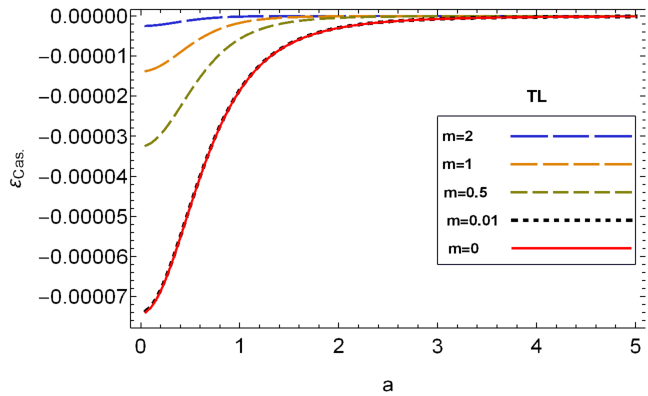}\includegraphics[width=8.5cm]{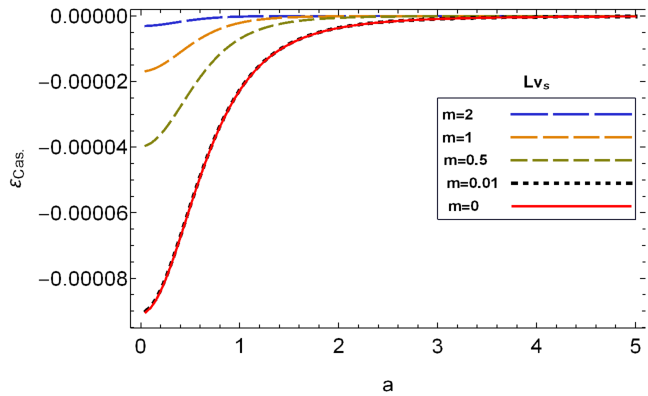}
\hspace{0.2cm}\includegraphics[width=8.5cm]{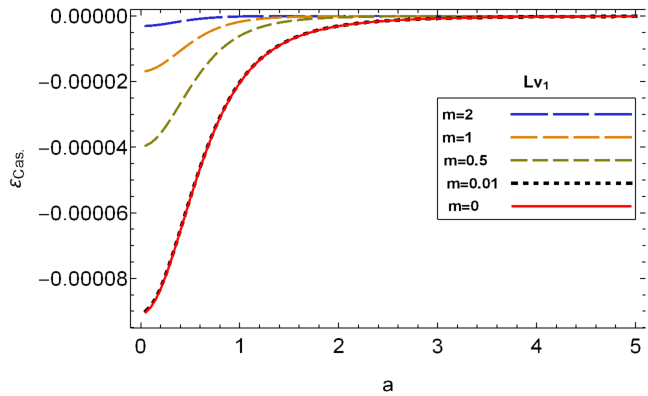}\includegraphics[width=8.5cm]{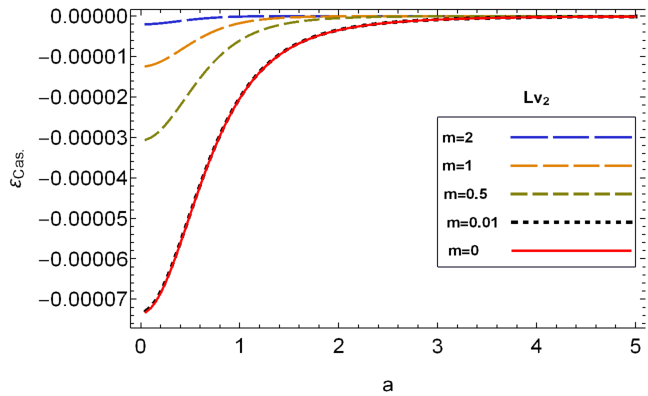}
\caption{\label{PLOT.RC.MASSIVE.MASSLESS} \small
  This figure illustrates the radiative correction to the Casimir energy density as a function of parameter $a$ in a three-dimensional space, for a range of mass values. Each subplot represents a different case of Lorentz violation, including TL, $Lv_s$, $Lv_1$, and $Lv_2$. It is evident from the trend that there is a prompt convergence from the massive to the massless cases. The divergence between the plots for massive cases with $m\leq0.01$ and the plot for the massless case is negligible. The parameters $\alpha$ and $h$ are held constant at $0.1$ and $1$, respectively, across all plots.}
\end{figure}
Figure (\ref{PLOT.RC.MASSIVE.MASSLESS}) was visually presented the radiative correction to the Casimir energy for both massive and massless scalar fields. It is worth mentioning that the convergence of the plotted curves for the massive fields, as the parameter $m$ tends towards zero, is seamlessly consistent with physical principles. This convergence signifies the agreement between the outcomes for the massive cases and the plot corresponding to the massless case. An intriguing aspect of this inquiry pertains to whether the sole contribution stemming from Lorentz symmetry breaking at the leading-order of the Casimir energy can nullify the radiative correction term of the Casimir energy within a system preserving Lorentz symmetry. To elucidate this, we initiated our investigation with a straightforward example, focusing on the leading-order Casimir energy density in the context of a massless scalar field subject to Lorentz violation via a time-like vector ($TL$) or Lorentz-violating scalar ($Lv_s$) field, while adhering to helical boundary conditions in three spatial dimensions:
\begin{eqnarray}\label{ZO.TL&LVs.LV.dim3}
    \mathcal{E}^{(0)}_{\mbox {\tiny Cas.}}=\frac{-\pi ^2}{90 a^4 \gamma ^2 \sqrt{1+s\alpha}}.
\end{eqnarray}
Expanding above expression in the limit $\alpha\to0$, we simply obtain
\begin{eqnarray}\label{ZO.TL&Lvs.LV.limit alpha to 0}
     \mathcal{E}^{(0)}_{\mbox {\tiny Cas.}}=-\frac{\pi ^2}{90a^4 \gamma ^2}+\frac{\pi ^2 \alpha  s}{180 a^4 \gamma ^2}+\mathcal{O}\left(\alpha ^2\right).
\end{eqnarray}
Furthermore, using Eq. (\ref{RC.massless.Cas.TL&Lvs}), the expression of radiative correction to the Casimir without Lorentz violation in $d=3$ is obtained as
\begin{eqnarray}\label{RC.Cas.final.TL&Lvs.dim3}
\mathcal{E}^{(R)}_{\mbox {\tiny Cas.}}=\frac{-\lambda }{1152 a^4 \gamma ^2}
\end{eqnarray}
The second term on the right-hand side of Eq. (\ref{ZO.TL&Lvs.LV.limit alpha to 0}) represents a significant portion of the deviation in the Casimir energy in the presence of Lorentz violation compared to its absence. This deviation value, when equalized with the radiative correction to the Casimir energy without Lorentz symmetry breaking as given in Eq. (\ref{RC.Cas.final.TL&Lvs.dim3}), establishes a relationship between the values of $\alpha$ and $\lambda$. This relationship is represented by $\alpha=\frac{5\lambda}{32\pi^2s}$. This relation imply the inherent contribution stemming from the TL/$Lv_s$ Lorentz symmetry breaking in the leading-order Casimir energy can offset the radiative correction term in the system devoid of Lorentz violation. Likewise, for the $Lv_1$/$Lv_2$ Lorentz violations, the corresponding value would be $\alpha=\frac{5\gamma\lambda}{32\pi^2(4\gamma c-8c-\gamma +4)}$. It is necessary to note that, the parameters $\alpha$ and $\lambda$ are inherently distinct quantities, and the established relationship does not imply a dependence between them. In Figure (\ref{3PLOTs}), we present a graphical comparison of the Casimir energy and its thermal and radiative corrections in three spatial dimensions. The figure clearly illustrates that as the dimensions increase, the energy value decreases. Furthermore, it demonstrates that the thermal correction consistently exhibits the opposite sign compared to the radiative correction. Moreover, the absolute value of the radiative correction is significantly smaller than that of the thermal correction.
\section{Conclusions}\label{sec:conclusion}
In this study, we have meticulously computed the Casimir energy for a Lorentz-violated massive/massless scalar field confined within helical boundary conditions across $d$ spatial dimensions. Notably, we have extended our analysis to encompass two critical corrections to this energy quantity: the thermal correction and the radiative correction. Both of these corrections to the Casimir energy when applying the helical boundary condition for the scalar field are novel. Furthermore, our determination of the radiative correction through a distinct renormalization program adds to the originality of our approach. Our chosen renormalization program, employing position-dependent counterterms, has been instrumental in deriving the first-order radiative correction to the vacuum energy. This approach not only incorporates the effects of boundary conditions or nontrivial backgrounds into the renormalization program but also ensures a self-consistent method for renormalizing the bare parameters of the Lagrangian. While similar renormalization programs have been discussed previously, our calculation extends its applicability to the Lorentz-violating scalar field. Our computed results regarding the radiative correction to the Casimir energy have exhibited convergence across all spatial dimensions, aligning with expected physical principles. Particularly noteworthy is the observation that in three spatial dimensions, the pure contribution from Lorentz symmetry breaking in the leading-order Casimir energy can nullify the radiative correction term ($\mathcal{O}(\lambda)$) in a system devoid of Lorentz violation. This phenomenon was illustrated for both massive and massless scalar fields within three spatial dimensions, and its generalizability to other space-time dimensions is suggested.
\begin{acknowledgments}
The author would like to thank the research office of Semnan Branch, Islamic Azad University for the financial support.
\end{acknowledgments}

\end{document}